\documentclass[useAMS,usedcolumn,usenatbib,usegraphicx]{mn2e}
\usepackage{times}

\usepackage{aas_macros}
\usepackage{textcmds}
\usepackage{multirow}
\usepackage{xcolor}
\usepackage{hyperref}
\usepackage{ctable}
\usepackage{amsmath,amssymb}
\newcommand\T{\rule{0pt}{2.6ex}}       % Top strut
\newcommand\B{\rule[-1.2ex]{0pt}{0pt}} % Bottom strut
\title[The chemistry of outflow cavity walls exposed]{The complex chemistry of outflow cavity walls exposed: the case of low-mass protostars}

\author[Maria N. Drozdovskaya et al.]{Maria~N.~Drozdovskaya$^{1}$\thanks{E-mail: drozdovskaya@strw.leidenuniv.nl}, Catherine~Walsh$^{1}$, Ruud~Visser$^{2}$, Daniel~Harsono$^{1,3}$\newauthor
 and Ewine~F.~van~Dishoeck$^{1,4}$\\
$^{1}$~Leiden Observatory, P.O. Box 9513, 2300 RA, Leiden, The Netherlands\\
$^{2}$~European Southern Observatory, Karl-Schwarzschild-Strasse 2, 85748 Garching, Germany\\
$^{3}$~SRON Netherlands Institute for Space Research, P.O. Box 800, 9700 AV Groningen, The Netherlands\\
$^{4}$~Max-Planck-Institut f\"{u}r Extraterrestrische Physik, Giessenbachstrasse 1, 85748 Garching, Germany}

\begin{document}

\date{Accepted 2015 May 20.  Received 2015 May 7; in original form 2015 March 18}

\pagerange{\pageref{firstpage}--\pageref{lastpage}} \pubyear{2015}

\maketitle

\label{firstpage}

\begin{abstract}
Complex organic molecules are ubiquitous companions of young low-mass protostars. Recent observations suggest that their emission stems, not only from the traditional hot corino, but also from offset positions. In this work, 2D physicochemical modelling of an envelope-cavity system is carried out. Wavelength-dependent radiative transfer calculations are performed and a comprehensive gas-grain chemical network is used to simulate the physical and chemical structure. The morphology of the system delineates three distinct regions: the cavity wall layer with time-dependent and species-variant enhancements; a torus rich in complex organic ices, but not reflected in gas-phase abundances; and the remaining outer envelope abundant in simpler solid and gaseous molecules. Strongly irradiated regions, such as the cavity wall layer, are subject to frequent photodissociation in the solid phase. Subsequent recombination of the photoproducts leads to frequent reactive desorption, causing gas-phase enhancements of several orders of magnitude. This mechanism remains to be quantified with laboratory experiments. Direct photodesorption is found to be relatively inefficient. If radicals are not produced directly in the icy mantle, the formation of complex organics is impeded. For efficiency, a sufficient number of FUV photons needs to penetrate the envelope; and elevated cool dust temperatures need to enable grain-surface radical mobility. As a result, a high stellar luminosity and a sufficiently wide cavity favor chemical complexity. Furthermore within this paradigm, complex organics are demonstrated to have unique lifetimes and be grouped into early (formaldehyde, ketene, methanol, formic acid, methyl formate, acetic acid, glycolaldehyde) and late (acetaldehyde, dimethyl ether, ethanol) species.
\end{abstract}


%%%%%%%%%%%%%%%%%%%%%%%%%%%%%%%%%%%%%%%%%%%%%%%%%%%%%%%%%%%%%%%%%%%%%%%%%%%%%%%
\section{Introduction}
\label{intro}
Young, forming stars emerge in regions seeded with diverse molecules. From highly abundant deuterated species in prestellar cores (see \citealt{CeccarelliPPVI} for a review) to complex organic molecules in hot cores and corinos (see \citealt{HerbstvD2009} for a review), chemical variety spans the entire lifetime of a protostar. Complex organic species are particularly alluring due to their potential astrobiological implications.

Complex organic compounds are loosely defined in both chemistry and astronomy, but typically mean large ($\geq 6$ atoms) carbon-containing species \citep{HerbstvD2009}. They were first observed in hot cores surrounding high-mass protostars (e.g., \citealt{Lovas1979, Blake1987, Kuan2014}), but have since also been detected in the environs of several low-mass counterparts (e.g., \citealt{vD1995, Cazaux2003, Bottinelli2004a, Bottinelli2004b, Jorgensen2012, Sakai2013, Maury2014}). The proximity of low-mass protostars facilitates spatially resolved studies. So far, single-dish telescopes have not revealed the origin of complex organics conclusively, but this is changing -- progress is being made with powerful interferometers, such as the Submillimeter Array (SMA), the Plateau de Bure Interferometer (PdBI) and the Atacama Large Millimeter/submillimeter Array (ALMA).

It is well established that hot corinos (defined as the inner warm zones where $T_{\text{dust}} \geq 100$~K) are chemically rich; however, several studies have shown that colder envelopes and positions with impinging outflows may also glow with emission from complex organic molecules. Formic acid (HCOOH), methyl formate (HCOOCH$_{3}$), acetaldehyde (CH$_{3}$CHO), methyl cyanide (CH$_{3}$CN) and ethanol (C$_{2}$H$_{5}$OH) were detected at a distance of $\sim 10^{4}$~AU from the L1157--mm low-mass protostar ($\sim 11$~L$_{\sun}$, \citealt{Sugimura2011}) -- towards the B1 position in the brightest carbon monoxide (CO) clump in the blue outflow lobe \citep{Arce2008, Codella2009, Sugimura2011, Codella2015}. The authors suggest that some species are already present on the grain surfaces and are subsequently liberated as gases by passing shocks, while others (e.g., methyl cyanide and acetaldehyde) form in the gas phase.

An outflow-affected position in the Perseus B1-b dust core (near a namefellow $\sim 3$~L$_{\sun}$ protostar) and another in the Serpens core (specifically, SMM4-W, separated by $~\sim 3000$~AU from SMM4, which has a luminosity of $\sim 5$~L$_{\sun}$) were studied by \citet{Oberg2010B1b} and \citet{Oberg2011Serpens}, respectively. The former did not show emission from complex organic species in contrast to the latter, which was bright in acetaldehyde and dimethyl ether (CH$_{3}$OCH$_{3}$). The envelopes of B1-b, SMM4, and SMM1 ($\sim 30$~L$_{\sun}$) contain lines from ketene (H$_{2}$CCO), formic acid, methyl formate, acetaldehyde, dimethyl ether, methyl mercaptan (CH$_{3}$SH), propynal (HCCCHO), methoxy (CH$_{3}$O) and potentially oxirane (CH$_{2}$OCH$_{2}$) \citep{Oberg2010B1b, Oberg2011Serpens, Cernicharo2012}. The envelope encompassing the most luminous target, SMM1, has the highest overall abundance of complex molecules. It was proposed that the ice content was revealed thanks to photodesorption, which may be efficient on larger scales due to the presence of outflow cavities, through which FUV photons can escape.

\citet{Jaber2014} presented data from The IRAS16293 Millimeter and Submillimeter Spectral Survey (TIMASS, \citealt{Caux2011}). Detections of ketene, methyl formate, acetaldehyde, dimethyl ether, methyl cyanide and formamide (NH$_{2}$CHO) in the cold ($\lesssim 50$~K) envelope (spanning $\sim 6000$~AU) surrounding IRAS16293-2422 A and B (combined $\sim 22$~L$_{\sun}$) were asserted. This is supported by earlier work on this source, which showed spatial segregation for various species (e.g., \citealt{vD1995, Ceccarelli2000, Schoier2002, Bisschop2008}).

\citet{Oberg2011Serpens}, based on their observations and laboratory experiments \citep{Oberg2009}, argued that complex organics form sequentially. While CO is frozen out, HCO-rich chemistry prevails, leading to `cold' species like methyl formate and acetaldehyde. Upon warm up, CH$_{3/2}$-rich molecules are formed, building `hot' species like dimethyl ether. However, observations of ketene, methyl formate, acetaldehyde, and dimethyl ether in the prestellar core L1689B \citep{Bacmann2012}, and formic acid towards the dark cloud L183 \citep{Requena-Torres2007}, argue against this sequential chemistry scenario due to the lack of elevated temperatures in these cold ($\sim 11-12$~K) sources.

A third prestellar core, L1544, was observed to show emission from ketene, formic acid, acetaldehyde, and propyne (CH$_{3}$CCH), but not from other species (such as methyl formate and dimethyl ether) found towards L1689B \citep{Vastel2014}. \citet{Cernicharo2012} secured a detection of the methoxy radical towards B1-b in Perseus and, based on laboratory data, argued that it is most likely produced in the gas phase from reactions between CH$_{3}$OH and OH - a reaction which has since been calculated to have a significantly larger rate coefficient at low temperatures than previously thought \citep{Shannon2013}. In summary, complex organic molecules are associated with outflow cavity walls, envelopes and even cold, dark cores, but it remains unclear when and how the various species form.

The complete chemistry of complex organics has been eluding astrochemists. Models have shown that gas-phase pathways contribute to the complex organic budget only under hot core-like conditions, when the densities and gas temperatures are high ($\sim 10^{7}$~cm$^{-3}$ and $>100$~K, e.g., \citealt{CharnleyTielensMillar1992, RodgersCharnley2001}). Instead, complexity could be built up in the ices. One solid-phase scenario is based on atom addition reactions \citep{TielensCharnley1997}. This has been experimentally proven for the sequential hydrogenation chain leading to methanol \citep{TielensHagen1982, Watanabe2004, Fuchs2009}. An alternative scenario is that of processed `zeroth generation' ices (i.e., simple ices like CO, methane (CH$_{4}$) and CH$_{3}$OH), which have been subjected to far-ultraviolet (FUV) irradiation, heating and/or cosmic rays (CRs, e.g., \citealt{Garrod2008}). Experimentally, an array of species is produced upon FUV irradiation of CH$_{3}$OH and CO mixtures \citep{Oberg2009}. Recent ideas include combinations of radical-radical and hydrogenation reactions \citep{Fedoseev2015}, and revisions of gas-phase reaction rates, potentially making gas-phase chemistry non-negligible at lower temperatures \citep{Balucani2015}. In addition, alternative mechanisms, like Eley--Rideal and complex-induced reactions \citep{Ruaud2015}, or CR-induced diffusion \citep{Reboussin2014}, have also been invoked to pave the way to complexity. However, efficient mechanisms that couple ices to the gas phase resulting in appreciable observed abundances remain challenging.

To obtain more insight, this paper offers physicochemical models of envelopes harboring low-mass protostars, including the impact of outflow cavities. Previously, \citet{Bruderer2009II, Bruderer2010III} modelled a high-mass young stellar object (YSO) with an outflow cavity with a focus on ions and diatomic hydrides. The authors pre-calculated chemical abundances with a gas-phase network at a given time, which saves computational time. Instead, here, a two-phase chemical network is used for time-dependent calculations on a cell-by-cell basis. \citet{Visser2012} simulated the low-mass analogue of \citet{Bruderer2009II, Bruderer2010III}, but focused on a thin warm/hot layer along the cavity wall, from which the far-infrared CO and water emission originate, rather than the larger cooler layers of the envelope that are central to this work.

In this work, the dust temperature and the stellar radiation field of the system are computed with RADMC -- a 2D wavelength-dependent radiative transfer module. A comprehensive gas-grain chemical model is employed in order to simulate the distribution of complex organic molecules across the envelope as a function of time. The major upgrade in this paper, in comparison to the setup of \citet{Visser2012}, is the inclusion of various grain-surface processes. Our models test the hypothesis that enhanced irradiation of the envelope, due to outflow cavities, stimulates the observed appearance of complex organic molecules in the gas phase in the colder offset positions from the protostar. The physicochemical models are described in Section~\ref{model}. The results are presented in Section~\ref{results} and discussed in Section~\ref{discussion}. The main findings are summarised in Section~\ref{conclusions}.

%%%%%%%%%%%%%%%%%%%%%%%%%%%%%%%%%%%%%%%%%%%%%%%%%%%%%%%%%%%%%%%%%%%%%%%%%%%%%%%
\section{Physicochemical model}
\label{model}

\ctable[
 width = 0.4\textwidth,
 caption = {Density distribution, outflow cavity, and stellar parameters (fiducial model setup)}.,
 label = tbl:pparams
 ]{@{\extracolsep{\fill}}llr}{}{
 \hline
 Parameter & Units & Value \T\B\\
 \hline
 $p$ &  & $-1.7$ \T\\
 $r_{\text{in}}$ & AU & $35.9~$ \\
 $n_{\text{H}_{2}; \text{ in}}$ & cm$^{-3}$ & $4.9 \times 10^{8}$ \\
 $a_{\text{cav}}$ & AU & $4.3 \times 10^{4}$ \\
 $b_{\text{cav}}$ & AU & $6.5 \times 10^{3}$ \\
 $\alpha\left( z=1000 \text{AU} \right)$ & deg & $109~$ \\
 $\alpha\left( z=10~000 \text{AU} \right)$ & deg & $45~$ \\
 $T_{*}$ & K & $10~000~$ \\
 $L_{*}$ & L$_{\sun}$ & $35.7$ \B\\
 \hline}

\subsection{Physical setup}
\label{physmod}

The goal of this work is to model the physical structure and the chemical composition of the envelope affected by outflow cavities on scales of $10~000$~AU. A static setup is adopted, representative of a low-mass embedded protostar (Class 0/I). The chemistry is then computed in time at each point under constant physical conditions.

The gas density distribution ($n_{\text{H}}$) sets the underlying dust distribution upon the assumption of a gas to dust mass ratio ($r_{\text{gd}}$). Once the dust distribution is coupled with a set of stellar variables, the temperature and radiation field distributions are defined. \citet{Jorgensen2002} and \citet{Kristensen2010} fitted dust continuum emission stemming from envelopes with the D\textsc{usty} code \citep{IvezicElitzur1997}. D\textsc{usty} models are spherically symmetric power law density distributions. The authors imposed the spectral energy distributions (SEDs) and continuum radial profiles at 450 and 850~$\mu$m from the JCMT as restrictions on these models to arrive at a set of best-fitting parameters. Those results were updated in \citet{Kristensen2012} based on new \textit{Herschel} fluxes. In this work, the density distribution derived with the D\textsc{usty} model of NGC 1333-IRAS2A is taken as a test case for low-mass protostars with strong emission from complex organics.

In the modelling work of \citet{Kristensen2010, Kristensen2012}, outflow cavities were neglected. Here, their presence is simulated by carving out ellipsoidal cavities from the D\textsc{usty} density distribution. This has previously been done in the work of \citet{Bruderer2009II, Bruderer2010III} and \citet{Visser2012} with two slightly different parameterisations. In this work, equation 1 from \citet{Visser2012} is used,
\begin{equation}
\label{Rcav}
 R _{\text{cav}}=b_{\text{cav}}\sqrt{1-\left( \frac{z}{a_{\text{cav}}}-1 \right)^{2}},
\end{equation}
where $\left( R_{\text{cav}}, z_{\text{cav}} \right)$ are the cylindrical coordinates of the cavity wall, and $a_{\text{cav}}$ and $b_{\text{cav}}$ are the semimajor and semiminor axes, respectively. Within the outflow cavity a comparatively low constant density is assumed,
\begin{equation}
\label{nHcav}
 n_{\text{H}}\left( \text{cavity} \right)=2 \times 10^{4} \text{cm}^{-3},
\end{equation}
and elsewhere,
\begin{equation}
\label{nHenv}
 n_{\text{H}}\left( \text{elsewhere} \right)= n_{\text{H}; \text{ in}} \times \left( \frac{r}{r_{\text{in}}} \right)^{p},
\end{equation}
where $r$ is the radial spherical coordinate, $p$ is the power index of the density profile, $r_{\text{in}}$ is the inner boundary of the envelope and $n_{\text{H}; \text{ in}}$ is the H nuclei number density at $r_{\text{in}}$. It is assumed that $n_{\text{H}}=2 \times n_{\text{H}_{2}}$ (at all positions). With the selected values of $a_{\text{cav}}$ and $b_{\text{cav}}$, the full opening angle of the outflow cavity is $109^{\circ}$ at $z=1000$~AU and $45^{\circ}$ at $z=10~000$~AU. \citet{Plunkett2013} tabulated the full opening angles for this source (table~4, although it is unclear at what distance from the source the values are derived), and find $30^{\circ}$ and $56^{\circ}$ for the two outflows in the blue lobe, and $32^{\circ}$ and $93^{\circ}$ for the two in the red lobe. These values are comparable to the full opening angles in this model. For other sources, wider angles have also been measured, e.g., $110^{\circ}$ for HH46/47 \citep{Velusamy2007}.

The amount of radiation stemming from a star depends on its blackbody temperature and the size of the emitting surface, i.e., on the stellar temperature ($T_{*}$) and radius ($R_{*}$). The bolometric luminosity of NGC 1333-IRAS2A is $35.7$~L$_{\sun}$ \citep{Kristensen2012, Karska2013}. Protostars are thought to have an ultraviolet (UV) excess due to accretion of material from the disc onto the star (e.g., \citealt{Spaans1995}). FUV photons are particularly efficient at photodissociating and photodesorbing molecules. Therefore in this work, a stellar temperature of $10~000$~K is adopted in order to account for the anticipated UV excess. From the definition of stellar luminosity ($L_{*}$), the following parameterization is used to obtain the stellar radius:
\begin{equation}
 R_{*}=\sqrt{\frac{L_{*}/\text{L}_{\sun}}{\left( T_{*}/\text{T}_{\sun} \right)^{4}}}\text{R}_{\sun}.
\end{equation}
Although parameters derived for NGC 1333-IRAS2A are used, our aim is not to model this particular source, but instead to build a template for a Class 0 protostar (see, e.g., fig.~13 in \citealt{Evans2009} for statistics per Class). The set of adopted parameters is summarized in Table~\ref{tbl:pparams}.

Once the dust density distribution and stellar properties are prescribed, \textsc{RADMC}\footnote[1]{\url{http://www.ita.uni-heidelberg.de/~dullemond/software/RADMC/index.shtml}} \citep{DullemondDominik2004} is employed to calculate the temperature and radiation field distributions. An $r_{\text{gd}}$ of $100$ is assumed and the dust mass density ($\rho_{\text{dust}}$) is given by:
\begin{equation}
 \rho_{\text{dust}}=0.5 \times n_{\text{H}} \times \mu \times m_{\text{p}}/r_{\text{gd}},
\end{equation}
where $m_{\text{p}}$ is the proton mass and $\mu$ is the mean molecular mass of the gas (taken to be $2.3$). The dust is assumed to be a mixture of carbonaceous material ($25$~per~cent) and silicates ($75$~per~cent). Opacity tables for bare grains covering a range of grain sizes from \citet{Pontoppidan2007a} are adopted. Modifications of the opacities to include icy mantles are expected to alter the dust temperatures ($T_{\text{dust}}$) by, at most, a few K (M.K. McClure, priv. comm.). Computations are carried out over frequencies in the $2.998 \times 10^{10} - 3.288 \times 10^{15}$~Hz ($0.09 - 1000$~$\mu$m) range, ensuring coverage of the dust's crystalline silicate features and the FUV range. The FUV radiation field ($F_{\text{FUV}}$) distribution is obtained via integration over the $912 - 2066$~$\text{\AA}$ ($6.0 - 13.6$~eV) range at every grid point. Isotropic scattering is included in the model. In this way, thanks to the curved cavity rims, direct irradiation of the envelope by the protostar is also simulated (as illustrated in fig.~5 of \citealt{Bruderer2009II}). Each setup was computed 10 times and the median $F_{\text{FUV}}$ was adopted for each grid cell, in order to minimize the numerical noise stemming from the Monte--Carlo nature of \textsc{RADMC}, while remaining computationally economic.

\subsection{Complex organic molecule chemistry}
\label{COMchem}

The evolving composition of the system is computed with a two-phase (gaseous and solid) chemistry code designed to solve for the abundances of species in time. The reaction rates depend on the physical conditions and the chemical network, as detailed in \citet{Walsh2014b, Walsh2014a}. The network has also been used previously in \citet{Drozdovskaya2014}, with only minor updates incorporated into the version used here. The chemical network is a compilation of the R\textsc{ate12} release of the UMIST Database for Astrochemsitry (UDfA\footnote[2]{\url{http://www.udfa.net}}, \citealt{McElroy2013}) and the Ohio State University (OSU) network \citep{Garrod2008}, with a total of $668$ species and $8764$ reactions.

The chemical model includes gas-phase two-body associations, recombination of cations on grain surfaces, adsorption, thermal desorption, CR-induced thermal desorption (i.e., spot heating), photodesorption, photodissociation and ionization in the gas and solid phases, grain-surface two-body reactions, and reactive desorption with an efficiency of $1$~per cent \citep{Garrod2007, VasyuninHerbst2013}. Two sources of FUV photons are accounted for: the protostar with the attenuated $F_{\text{FUV}}$ computed from \textsc{RADMC}, and the CRs, which produce FUV photons through excitation of H$_{2}$ (the most abundant molecule). Within the code, $F_{\text{FUV}}$ is converted to visual extinction ($A_{\text{V}}$) via
\begin{equation}
\label{Av}
 A_{\text{V}}=\tau/3.02,
\end{equation}
where $\tau$ is the UV extinction \citep{Bohlin1978} given by:
\begin{equation}
\label{tau}
 \tau=-\text{ln}\left( \frac{F_{\text{FUV}}}{\pi \times \int_{\text{FUV}}{B_{\lambda}\left(T_{*}\right) d\lambda} \times R_{*}^{2} / \left( R^{2}+z^{2} \right)} \right),
\end{equation}
where the denominator is the blackbody radiation for $T_{*}$ over the same FUV wavelength range as the numerator, with the inclusion of geometrical dilution, and where $\pi$ accounts for radiation stemming from one hemisphere towards a point in the envelope. A similar concept was applied in \citet{Visser2011} and \citet{Drozdovskaya2014}. A CR ionization rate ($\zeta_{\text{CR}}$) of $5 \times 10^{-17}$~s$^{-1}$ is used \citep{Dalgarno2006}, which is slightly higher than the traditional dark cloud value ($\zeta_{\text{ISM}}=1.3 \times 10^{-17}$~s$^{-1}$). FUV photons stemming from CR excitation of H$_{2}$ (with a flux of $10^{4}$~photons~cm$^{-2}$~s$^{-1}$ for $\zeta_{\text{ISM}}$, \citealt{Shen2004}) are produced internally, and are thus available in the most dense and shielded regions. The elevated $\zeta_{\text{CR}}$ favors chemical complexity, although recent work suggests that the ionization rate may be as much as $3$ orders of magnitude lower (e.g., \citealt{Padovani2013, Cleeves2013, Cleeves2015}). It is assumed that CRs penetrate the entire system uniformly, since even the maximum gas column density ($\sim1-2$~g~cm$^{-2}$) is too low for CR attenuation. Furthermore, magnetic fields are not modelled in this work; and stellar winds are expected to be important only in the innermost regions \citep{Cleeves2013}. External irradiation is not considered, since it strongly varies from source to source and is expected to affect only the very outermost envelope. H$_{2}$, CO and N$_{2}$ self- and mutual shielding is accounted for in the calculation of the respective photodissociation rates \citep{Visser2009photodis, Li2013}. The main difference between the chemical setup in \citet{Drozdovskaya2014} and the one here, is the use of photorates\footnote[3]{\url{http://home.strw.leidenuniv.nl/~ewine/photo/}} for a $10~000$~K rather than a $4000$~K blackbody \citep{vD2006, vanHemert2008}.

To decrease the computational weight of the chemistry module, all grains are assumed to be $0.1$ $\mu$m in radius ($a$), which is near the peak of typical dust populations. The number density of grain surface sites ($N_{\text{sites}}$) is $1.5\times10^{15}$ cm$^{-2}$ with a barrier thickness of $1$~$\text{\AA}$, under the assumption of a rectangular barrier \citep{HasegawaHerbstLeung1992}. Grains are considered to be either neutral or negatively charged, with a total, temporally and spatially constant number density ($x_{\text{grains}}$) relative to $n_{\text{H}}$. It is assumed that the gas temperatures equal the dust temperatures derived with \textsc{RADMC}. This should only break down at the lowest densities ($10^{3}-10^{4}$~cm$^{-3}$), e.g., within the outflow cavities themselves (where the chemistry is disregarded in this work) and in a thin layer along the cavity walls \citep{Visser2012}. Furthermore, gas temperature is of limited influence on the abundance of complex organics, because grain-surface associations and thermal desorption both depend solely on the dust temperature. Grain-surface reaction rates are computed assuming the Langmuir-Hinshelwood mechanism only and using the rate-equation method \citep{HasegawaHerbstLeung1992}.

Currently, in the chemical network, the formation of complex organic molecules relies primarily on associations of radicals upon FUV irradiation and the warm-up of icy mantles. At the lowest temperatures, hydrogenation is assisted by quantum tunneling of H and H$_{2}$ between grain surface sites, which becomes faster than classical thermal hopping. Quantum tunneling through reaction barriers is also allowed for these two species. The mobility of any species on the grain surface depends on its binding energy ($E_{\text{des}}$, also called its desorption energy). Here, the diffusion barrier ($E_{\text{diff}}$) is set by:
\begin{equation}
 E_{\text{diff}}=0.3 \times E_{\text{des}},
\end{equation}
as in \citet{HasegawaHerbstLeung1992}. This allows higher mobility at low temperatures ($\sim 15-20$~K) in comparison to the factor of $0.5$ used in \citet{GarrodHerbst2006}. Off-lattice kinetic Monte Carlo techniques suggest that the value is species-dependent, e.g., $0.31$ for CO and $0.39$ for CO$_{2}$ have been determined on the surface of crystalline water ice \citep{KarssemeijerCuppen2014} -- values which are in line with that assumed in this work. The set of adopted $E_{\text{des}}$ values is the one supplied with R\textsc{ate12}, with the exception of water, for which a higher value of $5773$~K for pure water ice is used instead \citep{Fraser2001}. These values can vary significantly depending on the surface composition, which may vary with the formation sequence of icy mantles and eventual segregation of solid species within the mantle.

FUV photons are responsible for photodissociating molecules and producing radicals, which may go on to make larger species. Radicals can be formed directly in the ice or in the gas phase, followed by subsequent adsorption or gas-phase reactions. However, FUV photolysis is a delicate balance. A strong FUV radiation field breaks molecules apart and favorably produces photofragments (like CH$_{3}$ and C$_{2}$H$_{5}$). In this work, an elevated $\zeta_{\text{CR}}$ favors radical production in strongly attenuated regions ($A_{\text{V}} \gtrsim 5$~mag) by CR-induced FUV photons (e.g., \citealt{Gerakines1996, Palumbo2008}), which has been seen to enhance complexity in experiments \citep{Maity2015}. Elsewhere, the stellar FUV field dominates. Other mechanisms for producing radicals have been postulated in the literature and supported by experiments, e.g., low-energy electron radiolysis \citep{Boamah2014}, but they are not taken into account in this work.

The assumed grain properties are highly simplified in this chemical setup. Grain sizes and their distribution change with time, although theory suggests that this is most prominent for protoplanetary discs (see \citealt{TestiPPVI} for a review; however there is also accumulating observational evidence for grain growth in the embedded phase, e.g., \citealt{Jorgensen2009, Ricci2010, Lommen2010, Pagani2010, Miotello2014}). The grains are also not necessarily spherical, and their surfaces may be highly irregular. All these parameters will change the surface area available for grain-surface chemistry and the number of binding sites that are available per unit volume. Under the current assumptions, the key parameter is the total density of sites ($n_{\text{sites}}$), given by:
\begin{equation}
n_{\text{sites}}=4 \pi \times a^{2} \times N_{\text{sites}} \times x_{\text{grains}} \times n_{\text{H}}.
\end{equation}
Here, $x_{\text{grains}}=2.2 \times 10^{-12}$ is assumed. Decreasing this number by a factor of $\sim 2$, for example, results in abundances of gaseous and solid species changing by a factor of $\sim 2-3$ non-linearly.

Finally, time is also an important parameter. Grain-surface radical-radical association reaction rates are highly temperature-dependent and may require a long time to build up appreciable product abundances. Thermal desorption, on the other hand, can be very rapid if the dust temperature is sufficiently high. Radicals may thus be lost to the gas phase before a reaction can take place. In this work, the chemical evolution of the system is computed up to $3 \times 10^{6}$~yr, which corresponds to the upper limits on the age of embedded sources (see table~1 in \citealt{DunhamPPVI}).

%%%%%%%%%%%%%%%%%%%%%%%%%%%%%%%%%%%%%%%%%%%%%%%%%%%%%%%%%%%%%%%%%%%%%%%%%%%%%%%
\section{Results}
\label{results}

\begin{figure}
 \centering
 \includegraphics[keepaspectratio]{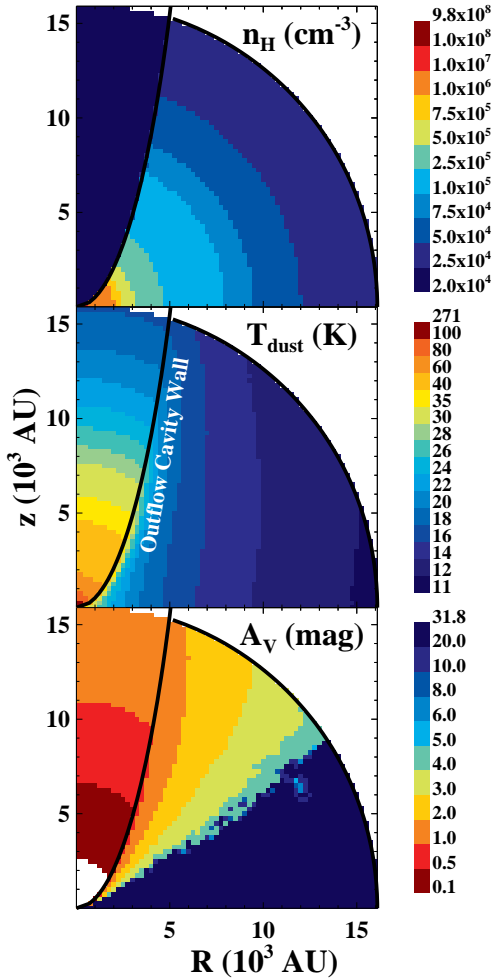}
 \caption{The physical structure of the system, from top to bottom: the gas density, $n_{\text{H}}$ (cm$^{-3}$), the dust temperature, $T_{\text{dust}}$ (K) and the visual extinction, $A_{\text{V}}$ (mag). The outflow cavity wall is shown with a black curve. Data points only within a fixed radius ($16~100$~AU) of the protostar are considered.}
 \label{fgr:phys}
\end{figure}

\ctable[
 width = 0.48\textwidth,
 caption = Abundances at the end of the prestellar core phase\tmark.,
 label = tbl:mabun
 ]{@{\extracolsep{\fill}}llll}{
 \tnote{The physical conditions of the prestellar core phase are assumed to be $n_{\text{H}}=4 \times 10^{4}$~cm$^{-3}$, $T_{\text{dust}}=10$~K, $F_{\text{FUV}}=0$~erg cm$^{-2}$ s$^{-1}$ and $t=3 \times 10^{5}$~yr.}
 }{
 \hline
 Species & Name & $n \left( \text{X}_{\text{gas}} \right) / n_{\text{H}}$ & $n \left( \text{X}_{\text{ice}} \right) / n_{\text{H}}$ \T\B\\
 \hline
 CO & carbon monoxide & $3.5\times10^{-5}$ & $5.6\times10^{-5}$ \T\\
 H$_{2}$O & water & $5.7\times10^{-8}$ & $2.0\times10^{-4}$ \\
 H$_{2}$CO & formaldehyde & $6.0\times10^{-8}$ & $1.4\times10^{-5}$ \\
 CH$_{2}$CO & ketene & $6.7\times10^{-9}$ & $1.7\times10^{-8}$ \\
 CH$_{3}$OH & methanol & $1.3\times10^{-10}$ & $9.2\times10^{-6}$ \\
 HCOOH & formic acid & $1.1\times10^{-10}$ & $4.4\times10^{-10}$ \\
 HCOOCH$_{3}$ & methyl formate & $3.2\times10^{-15}$ & $4.4\times10^{-15}$ \\
 CH$_{3}$CHO & acetaldehyde & $2.1\times10^{-9}$ & $1.1\times10^{-9}$ \\
 CH$_{3}$OCH$_{3}$ & dimethyl ether & $3.1\times10^{-15}$ & $3.2\times10^{-12}$ \\
 C$_{2}$H$_{5}$OH & ethanol & $5.1\times10^{-12}$ & $4.3\times10^{-12}$ \\
 CH$_{3}$COOH & acetic acid & $2.7\times10^{-21}$ & $5.5\times10^{-18}$ \\
 HOCH$_{2}$CHO & glycolaldehyde & $8.6\times10^{-21}$ & $2.4\times10^{-17}$ \B\\
 \hline}

\begin{figure}
 \centering
 \includegraphics[keepaspectratio]{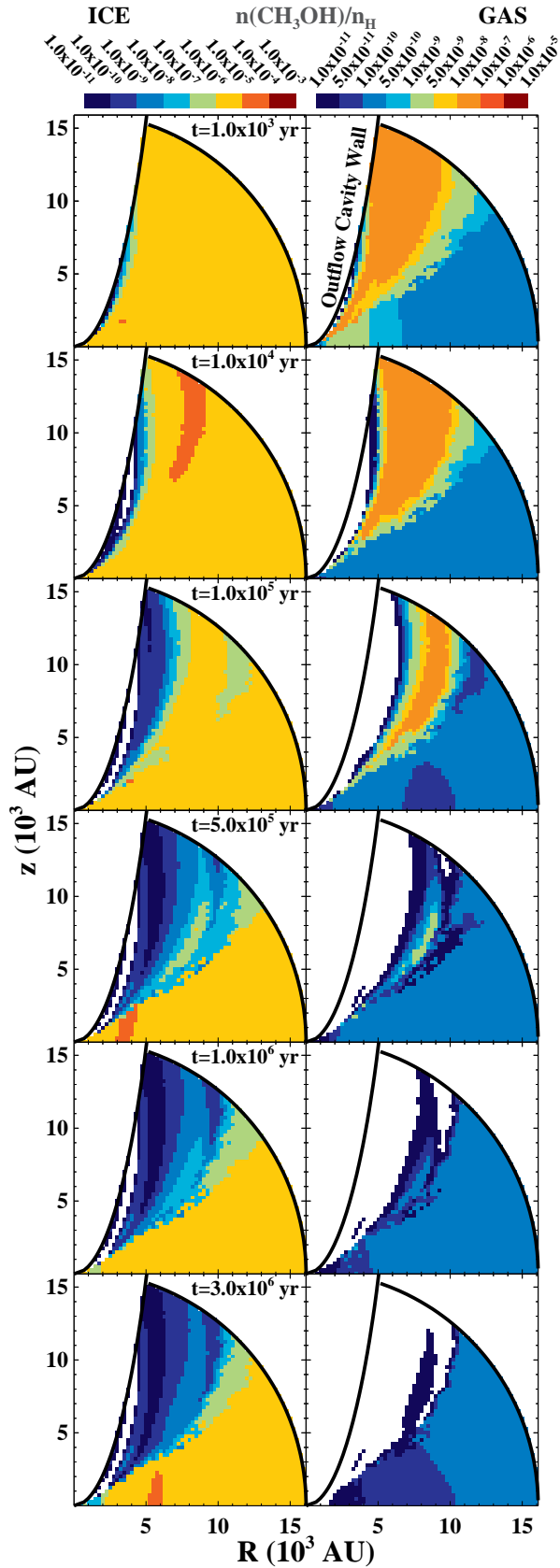}
 \caption{The abundance of methanol in the solid (left column) and gas (right column) phases at six different time steps across the envelope-cavity system. The outflow cavity wall is shown with a black curve. White cells correspond to either being outside of the area being considered or to having values outside of the range of the colour bar. The range of the gas colour bar is different from the range of that of the ice.}
 \label{fgr:CH3OH}
\end{figure}

\begin{figure}
 \centering
 \includegraphics[keepaspectratio]{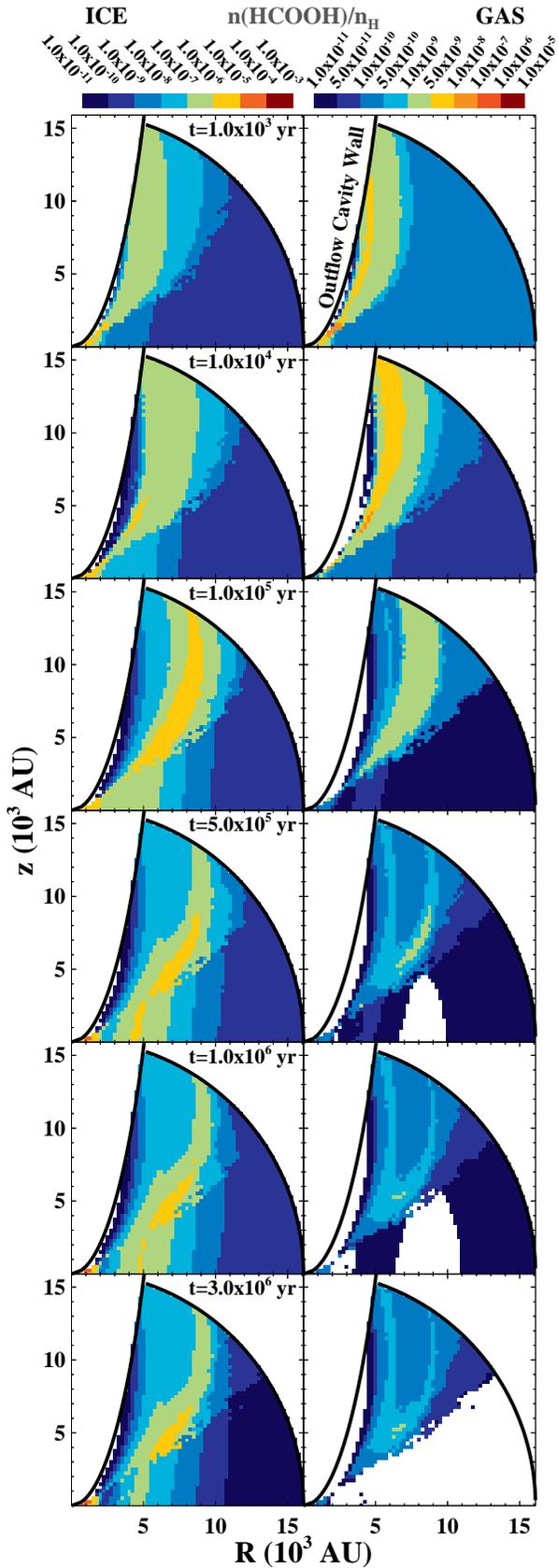}
 \caption{Same as Fig.~\ref{fgr:CH3OH}, but for formic acid.}
 \label{fgr:HCOOH}
\end{figure}

\begin{figure}
 \centering
 \includegraphics[keepaspectratio]{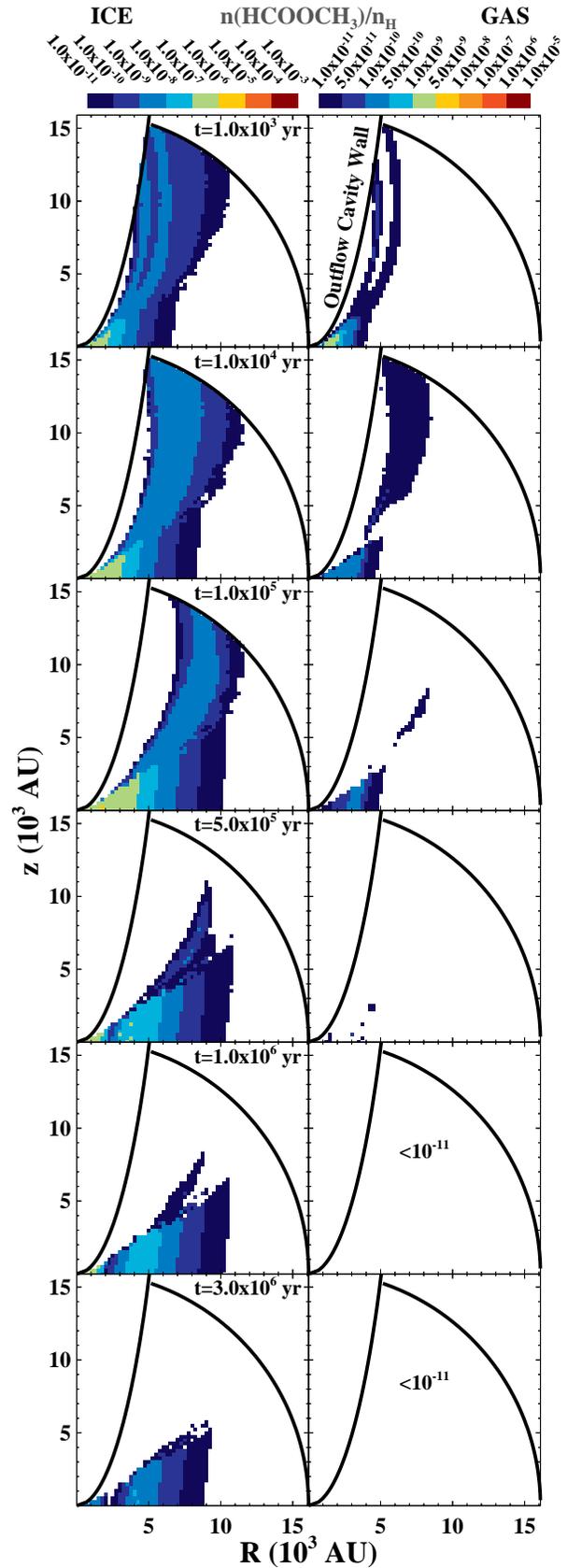}
 \caption{Same as Fig.~\ref{fgr:CH3OH}, but for methyl formate.}
 \label{fgr:HCOOCH3}
\end{figure}

\begin{figure}
 \centering
 \includegraphics[keepaspectratio]{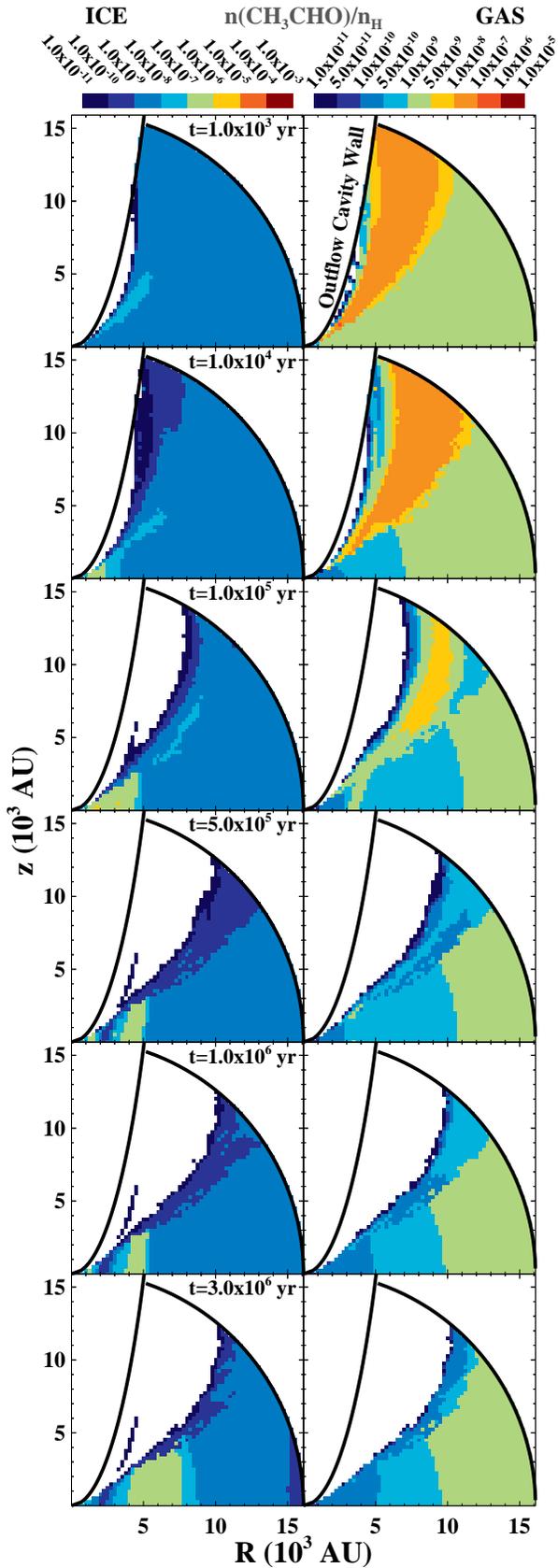}
 \caption{Same as Fig.~\ref{fgr:CH3OH}, but for acetaldehyde.}
 \label{fgr:CH3CHO}
\end{figure}

\begin{figure}
 \centering
 \includegraphics[keepaspectratio]{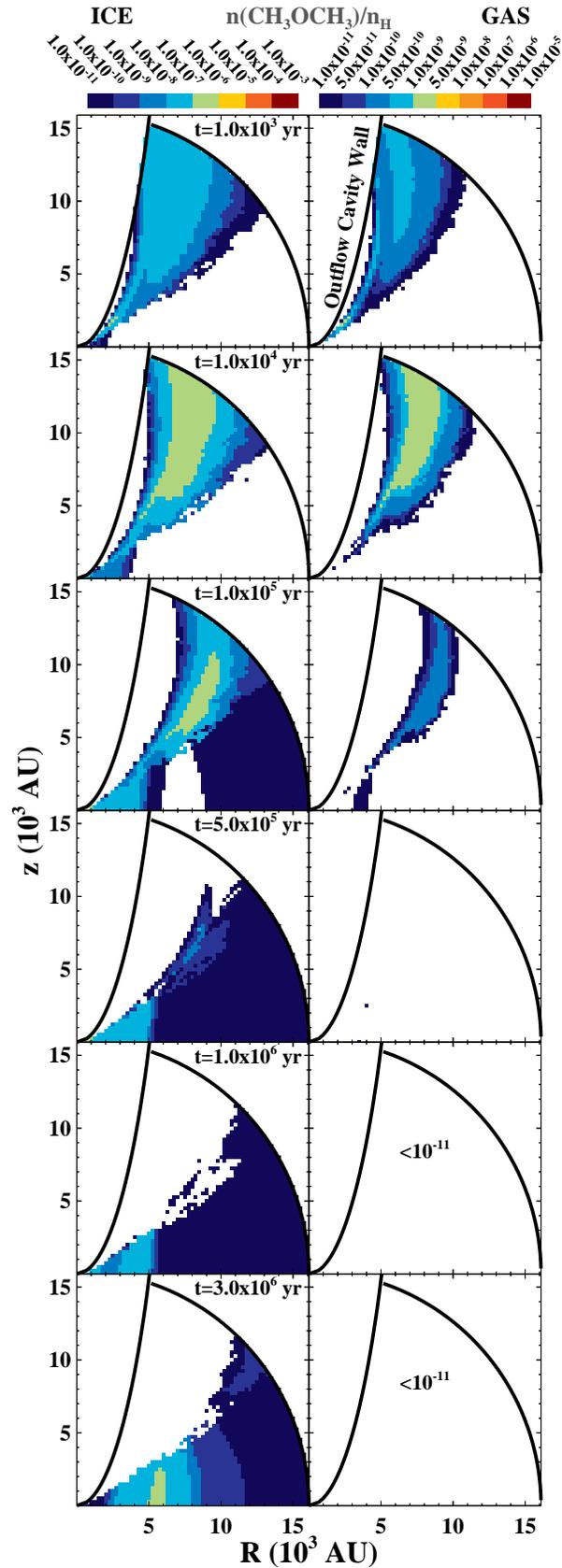}
 \caption{Same as Fig.~\ref{fgr:CH3OH}, but for dimethyl ether.}
 \label{fgr:CH3OCH3}
\end{figure}

\subsection{Physical structure}

The physical structure of the system is displayed in Fig.~\ref{fgr:phys}. In the top panel, the number density of H nuclei follows the adopted power law throughout the envelope, $n_{\text{H}} \propto r^{-1.7}$ (equation~\ref{nHenv} and Table~\ref{tbl:pparams}). The outflow cavity has a fixed low density of $2 \times 10^{4}$~cm$^{-3}$ (equation~\ref{nHcav}). The ellipsoidal cavity wall, as specified by equation~\ref{Rcav} for the parameters given in Table~\ref{tbl:pparams}, separates the zone carved out by the outflow from the envelope. The middle panel shows the dust temperature as obtained with \textsc{RADMC}. The cavity is the warmest region with $T_{\text{dust}} > 100$~K within the inner $\sim 300$~AU of the protostar, as a result of its low density and thus low extinction. The envelope is predominantly cold, $T_{\text{dust}} \lesssim 25$~K. Only a thin strip of material along the cavity wall (for $z \lesssim 8000$~AU) attains lukewarm conditions, $\sim 30-40$~K. Along the midplane, the $T_{\text{dust}} = 100$~K threshold is exceeded at $R \approx 38$~AU. This, in combination with our exclusion of the cavity itself, implies that the traditional hot corino is not treated in this work. The bottom panel displays the visual extinction, which is computed with equations~\ref{Av} and~\ref{tau} based on the stellar FUV radiation field calculated with \textsc{RADMC}. Again due to its low density, the cavity is subject to the strongest irradiation, $A_{\text{V}} < 2$~mag. The envelope zones closest to the radial axis are fully shielded from stellar FUV photons, $A_{\text{V}} > 10$~mag, thanks to the large dust column density separating them from the protostar. The rest of the envelope is subject to weak FUV irradiation, $A_{\text{V}} \lesssim 3$~mag.

The inner $35.9$~AU around the protostar are excluded from this setup (Table~\ref{tbl:pparams}), where the physical conditions need to be adjusted for the presence of a protoplanetary disc. The \textsc{RADMC} calculations are performed on a much finer grid than that used for plotting in Fig.~\ref{fgr:phys}, which is the adopted grid for time-consuming chemical computations ($4290$ points in total). All cells are treated individually in the subsequent chemical calculations, i.e., as in a static model (in contrast with a dynamic evolutionary setup, as in \citealt{Drozdovskaya2014}). This maximizes the time spent in each temperature regime.

\subsection{Initial abundances}

The output of a single-point model of the prestellar core phase under constant physical conditions is assumed to be representative of the envelope material and therefore, is adopted as the set of initial molecular abundances. This scheme was also followed by \citet{Visser2011} and \citet{Drozdovskaya2014}. The assumed physical parameters of the prestellar core are $n_{\text{H}}=4 \times 10^{4}$~cm$^{-3}$, $T_{\text{dust}}=10$~K, negligible stellar FUV flux and an age of $3 \times 10^{5}$~yr. \citet{AndrePPVI} (in their section~3.3) estimated that the average lifetime of a starless core with a typical density of $\sim 10^{4}$~cm$^{-3}$ is $\sim 10^{6}$~yr. A selection of the obtained abundances is provided in Table~\ref{tbl:mabun}. Dark core observations of L1689B \citep{Bacmann2012}, L1544 \citep{Vastel2014} and observations towards the `core' position in B1-b of \citet{Oberg2010} (and data from \citealt{Cernicharo2012}), derive column densities on the order of $\sim 10^{12}$~cm$^{-2}$ for ketene, formic acid, methyl formate, acetaldehyde and dimethyl ether. If a typical H$_{2}$ column density of $\sim 10^{23}$~cm$^{-2}$ is used, then the observed abundances of these species in prestellar cores are $\sim 10^{-11}$ within an order of magnitude. A comparison with the values in Table~\ref{tbl:mabun} reveals that from the aforementioned list only gas-phase methyl formate and dimethyl ether are underestimated in this prestellar core model. On the other hand, the modelled abundances of simpler ices, like water, carbon monoxide and methanol, are close to the observed ice abundances against background stars (e.g., \citealt{Knez2005, Boogert2011, Boogert2013, Boogert2015}).

\subsection{Abundance maps}

Figs~\ref{fgr:CH3OH}-\ref{fgr:CH3OCH3} show the abundance maps of five different molecules at six different time steps. The left six panels display the solid phase and the right six panels show the gas phase. For completeness, the abundance maps of five additional molecules (water, formaldehyde, ketene, ethanol, acetic acid) are provided in Appendix~\ref{addmaps} (and glycolaldehyde is discussed in Section~\ref{glysection}).

\subsubsection{Methanol}

Fig.~\ref{fgr:CH3OH} shows that the methanol ice abundance remains predominantly unchanged from its initial abundance ($9.2\times10^{-6}$, Table~\ref{tbl:mabun}) for the first $10^{3}$~yr. With time, the abundance drops outwards from the cavity wall and eventually methanol ice is abundant ($\sim 10^{-6}$) only in the lower half of the envelope. This is the effect of photodissociation of methanol ice by stellar FUV photons (recall the $A_{\text{V}}$ map from Fig.~\ref{fgr:phys}). At $10^{4}$~yr, the upper cavity wall layer displays a small enhancement (within an order of magnitude). Stellar FUV photons do not only destroy methanol ice, but they also produce CH$_{3}$ and OH radicals by photodissociating other species (like H$_{2}$O ice), providing an additional formation channel for methanol ice (besides sequential hydrogenation of CO, which is efficient across the entire envelope). This enhancement along the wall survives until $10^{5}$~yr, when the more distant zone (at $z \sim 12~000$~AU and $R \sim 14~000$~AU) starts to show a lower abundance ($\sim 10^{-7}$) by an order of magnitude, indicating favouritism towards destruction in that region.

Another methanol ice-rich zone is formed around $z \sim 500$~AU and $R \sim 5000$~AU. It will be shown in subsequent maps (Fig.~\ref{fgr:HCOOH}-\ref{fgr:CH3OCH3}) that this small area is rich in all complex organic ices, and for that reason this area is referred to as the `complex organic molecule torus'. Methanol, being a key parent species, is also abundant here. As it is converted to larger species, the enhancement is subdued at $10^{6}$~yr. At longer time-scales, $\sim 3 \times 10^{6}$~yr, CR-induced FUV photons start to dominate and photodissociate larger molecules to remake methanol, leading to the regaining of the enhancement and a resetting of the ice chemical complexity.

The distribution of gas-phase methanol is closely related to the abundance of methanol ice.  In the regions furthest away from the cavity, where methanol ice is abundant ($\sim 10^{-6}$) for all times, the abundance of methanol gas is low ($\lesssim 5 \times 10^{-10}$). The entire envelope is colder than the thermal desorption temperature of methanol ($\sim 90$~K), and there are no efficient gas-phase routes to methanol, therefore the amount of gaseous methanol relies purely on non-thermal desorption. In the furthest regions, extinction is high, so photodesorption with stellar FUV photons is insignificant. Gas-phase methanol is obtained through weak photodesorption by CR-induced FUV photons and reactive desorption stemming from hydrogenation reactions with CO. The cavity walls, on the other hand, light up in gaseous methanol for the first $\sim 10^{5}$~yr. This corresponds to the zone showing the methanol ice enhancement. There, enough stellar FUV photons penetrate to allow more efficient photodesorption, but more importantly, to dissociate methanol ice. Upon the subsequently recombination of the photoproducts, reactive desorption is assumed to occur in $1$~per~cent of such reactions, enhancing the gas-phase methanol abundance.

\subsubsection{Formic acid}

Formic acid has an initial abundance of $\sim 10^{-10}$ in both phases. There are several key grain-surface pathways leading to its formation, specifically hydrogenation (H~+~COOH) and associations of OH either with H$_{2}$CO or HCO. Solid formic acid is clearly enhanced along the cavity wall at all time steps in Fig.~\ref{fgr:HCOOH}. This occurs due to elevated radical-radical association rates thanks to stellar FUV photons, as was the case with methanol ice. The enhancement moves further away from the cavity rim with time with the separation between the two being largest ($\sim 4000$~AU) at highest $z$ values ($\sim 10~000$~AU), which is a result of the formation-destruction balance shifting. The zone with the highest abundances ($>10^{-6}$) is maximized at $10^{5}$~yr. Gas-phase formic acid illuminates the cavity wall primarily at the earlier times ($\leq 10^{5}$~yr). Its abundance peaks ($>5 \times 10^{-9}$)  at $10^{4}$~yr, before that of the ice. This corresponds to the time when radical-radical chemistry is most active. On the longest time-scales, CR-induced FUV photons destroy formic acid in both phases.

\subsubsection{Methyl formate}

Fig.~\ref{fgr:HCOOCH3} shows the distribution of methyl formate. Its ice reaches abundances of $\sim 10^{-9}$ along the cavity wall, which is a large increase from its initial abundance of $\sim 10^{-15}$. Peak abundances $> 10^{-7}$ are reached at $10^{5}$~yr in the complex organic molecule torus. Gaseous methyl formate exceeds the $\sim 10^{-11}$ abundance level at early times along the cavity wall, but mostly remains at a low abundance ($<10^{-11}$). The enhancement of the ice in the complex organic molecule torus is not reflected in the gas-phase abundance, because at its location, there are no efficient routes to liberate the icy mantle species. This was also the case with methanol and formic acid in Figs~\ref{fgr:CH3OH} and~\ref{fgr:HCOOH}, respectively. The formation of methyl formate predominantly occurs either via the association of methoxy with HCO or with CO followed by hydrogenation.

\subsubsection{Acetaldehyde}

Acetaldehyde ice remains at its initial abundance of $\sim 10^{-9}$ throughout the envelope. With time, photodissociation by stellar FUV photons depletes it in the vicinity of the cavity wall (Fig.~\ref{fgr:CH3CHO}). An appreciable enhancement is built up in the torus with an abundance of $\sim 10^{-7}-10^{-6}$. Gaseous acetaldehyde appears to be highly decoupled from its ice due to a strong enhancement ($>10^{-8}$) along the wall for $\lesssim 10^{5}$~yr, which is not seen in the solid phase. This feature appears due to a single gas-phase reaction, namely O~+~C$_{2}$H$_{5}$ (with a rate coefficient of $1.33 \times 10^{-10}$~cm$^{3}$~s$^{-1}$ upon extrapolation from \citealt{TsangHampson1986}). Atomic O is highest in abundance towards zones with the stronger irradiation, i.e., towards the cavity, while C$_{2}$H$_{5}$ is most abundant towards zones with the least irradiation, i.e., towards the radial axis. This results that in the `overlap region', the cavity wall layer, the reaction is very efficient at producing acetaldehyde. Grain-surface production of acetaldehyde proceeds via CH$_{3}$~+~HCO and CH$_{3}$~+~CO followed by a hydrogenation.

\subsubsection{Dimethyl ether}

For the first $\sim 10^{5}$~yr, dimethyl ether strongly highlights the cavity wall with solid and gaseous abundances higher than $10^{-7}$ and $10^{-9}$, respectively (Fig.~\ref{fgr:CH3OCH3}). This is the result of active grain-surface chemistry and concomitant reactive desorption. At longer times ($>10^{5}$~yr), the enhancement is destroyed due to the domination of photodissociation by stellar FUV photons. However, at those times dimethyl ether ice is built up in the torus at an abundance of $10^{-7}$, which is hidden from the gas phase for the above-stated reasons. The only grain-surface pathway towards dimethyl ether included in the network is the association of CH$_{3}$ with methoxy.

\subsection{Parameter study}

\begin{figure}
 \centering
 \includegraphics[keepaspectratio]{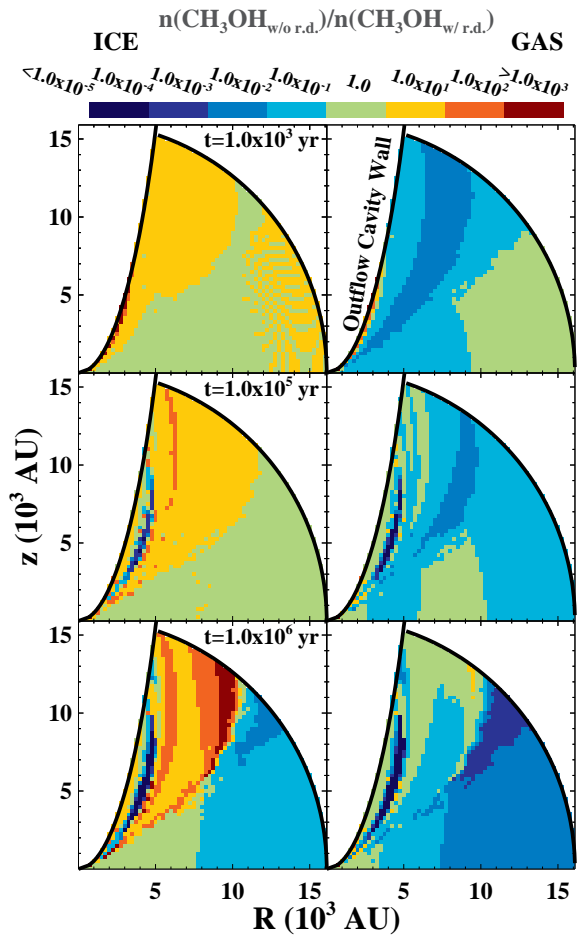}
 \caption{The ratio of the methanol abundance with reactive desorption switched off to that with an efficiency of $1$~per~cent. Three time steps are shown with the solid-phase and the gas-phase ratios in the left and right columns, respectively. The outflow cavity wall is shown with a black curve. A certain colour corresponds to a range of values between the two labeled bounds, e.g., yellow corresponds to any value between $1$ and $10$.}
 \label{fgr:RD}
\end{figure}

\begin{figure}
 \centering
 \includegraphics[keepaspectratio]{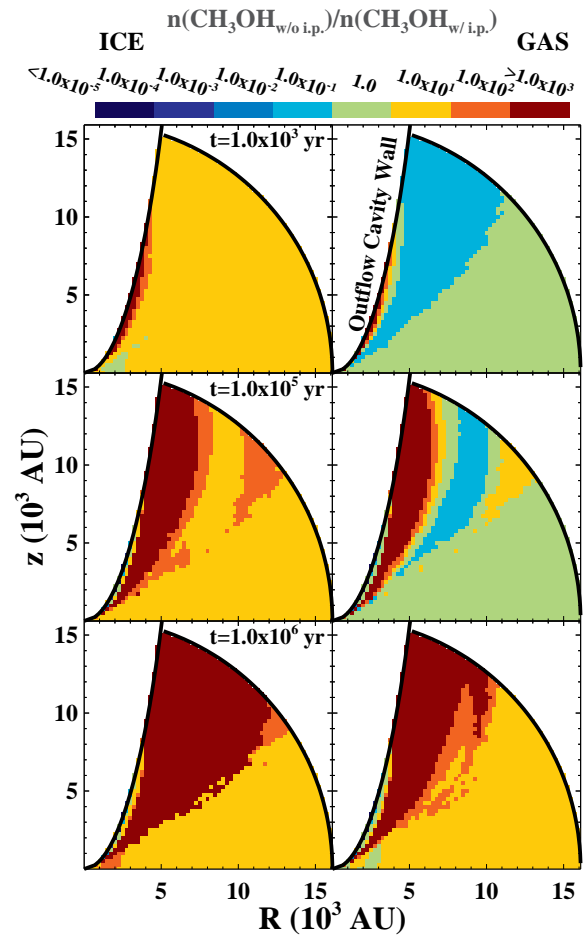}
 \caption{The ratio of the methanol abundance with ice photodissociation (by stellar and CR-induced FUV photons) switched off to that with it on. Three time steps are shown with the solid-phase and the gas-phase ratios in the left and right columns, respectively. The outflow cavity wall is shown with a black curve.}
 \label{fgr:IP}
\end{figure}

\subsubsection{Reactive desorption}

During the above analysis of abundance maps, the importance of reactive desorption is readily seen, even with a mere $1$~per~cent efficiency. This has also been reported in previous publications. \citet{VasyuninHerbst2013} saw variations of several orders of magnitude in gas-phase abundances and differences of $10$~per~cent in ice abundances for a dark core model ($T=10$~K, $n_{\text{H}}=10^{5}$~cm$^{-3}$, $A_{\text{V}}=10$~mag, $t=10^{5}-10^{6}$~yr) upon varying reactive desorption efficiency from $0$ to $10$~per~cent. \citet{Wakelam2014} noted that gas-phase water abundances can change by a factor of $10$ upon exclusion of reactive desorption (for various single point models with $T=15-30$~K, $n_{\text{H}}=2 \times 10^{4}-2 \times 10^{5}$~cm$^{-3}$, $A_{\text{V}}=2-4$~mag, $t=10^{4}-10^{6}$~yr). Experiments constraining this process are scarce. Recently, \citet{Dulieu2013} and \citet{Minissale2014} suggested that the efficiency of reactive desorption can vary greatly per reaction and per surface, from very efficient on bare grains to completely inefficient as soon as the first monolayer is built up.

Fig.~\ref{fgr:RD} shows the ratio of the methanol abundance with reactive desorption switched off to that with an efficiency of $1$~per~cent, i.e., comparing with Fig.~\ref{fgr:CH3OH}. Gas-phase abundances decrease over almost the entire envelope for all times once reactive desorption is excluded. The decrease is $1-3$ orders of magnitude for the largest portions of the envelope, and is explained by the exclusion of the most efficient mechanism for the population of the gas phase by the methanol-rich ices. In strongly irradiated areas, $A_{\text{V}}<3$~mag, the abundance of solid methanol is as much as $3$ orders of magnitude higher when reactive desorption is not included, which is an ice destruction mechanism. However, in zones where CR-induced FUV photons dominate, $A_{\text{V}}>3$~mag, there is less solid methanol when reactive desorption is excluded. Reactive desorption elevates the gas-phase abundance not only of methanol and complex organic molecules, but also of simpler species, like CH$_{4}$. Gas-phase ion-molecule reactions then lead to an enhancement of gas-phase radicals, most notably CH$_{3}$. Higher gaseous radical abundances imply greater availability of these species on the grains (transiently) as well, thus allowing more efficient formation of methanol via the OH~$+$~CH$_{3}$ route.

Acetaldehyde and dimethyl ether follow a trend very similar to that seen with methanol in Fig.~\ref{fgr:RD}. Formic acid and methyl formate differ by not showing a decrease in the zone with high extinction when reactive desorption is switched off. This is linked to them being less reliant on the availability of CH$_{3}$.

\subsubsection{Ice photodissociation}

Another poorly constrained process is the photodissociation of solid species. UV photons can penetrate as deep as $100$ monolayers into the icy mantle; however, what happens with the photofragments thereafter remains unclear. They may recombine immediately or diffuse away through the ice, if they have sufficient translational energy after photodissociation (e.g., \citealt{AnderssonvD2008}). In the current setup, it is assumed that the entire mantle may be dissociated; however, by making all those radicals equally available for further grain-surface reactions, the chemistry of the solid phase is potentially overestimated, because the diffusion rates within the bulk ice mantle are likely slower than those across the surface.

To test the significance of ice photodissociation by stellar and CR-induced FUV photons, a control simulation is carried out with those processes turned off. Photodesorption and reactive desorption are still included. Fig.~\ref{fgr:IP} shows the ratio of the methanol abundance with ice photodissociation excluded to that with it included, i.e., again comparing with Fig.~\ref{fgr:CH3OH}. The amount of methanol ice is increased roughly by an order of magnitude in zones with the highest extinction ($A_{\text{V}}>3$), and in the cavity wall by $3$ orders of magnitude. This corresponds to where photodissociation is dominated by stellar UV photons and is the primary ice destruction pathway. For $t \leq 10^{5}$~yr, the gas-phase methanol abundance is reduced, since less ice photodissociation implies less radical recombinations, and thus less gas-phase methanol due to reactive desorption. However, at later time steps and in the cavity walls, a much larger abundance of ice leads to more gaseous methanol via photodesorption, CR-induced thermal desorption, and reactive desorption associated with slow formation pathways. At the final time step, the gas-phase methanol abundance starts to decrease in the most shielded zone, since CR-induced FUV photons are still photodissociating in the gas-phase, but efficient mechanisms (in particular, reactive desorption, in comparison to the fiducial setup) to replenish the gas from the ice are absent.

Other complex organics, i.e., formic acid, methyl formate, acetaldehyde and dimethyl ether, show a decrease by several orders of magnitude in both phases once ice photodissociation is switched off. In this control simulation, the radical production is limited to the gas phase, while key radical sources, such as methanol, are most abundant in the solid phase. The reduction of radicals drastically obstructs efficient formation of complex organic molecules. Only along the cavity wall, are the abundances higher than fiducially. Ice photodissociation dominates there and in the standard case efficiently destroys complex organics.

\subsubsection{An additional route to glycolaldehyde}
\label{glysection}

\begin{figure*}
 \centering
 \includegraphics[keepaspectratio]{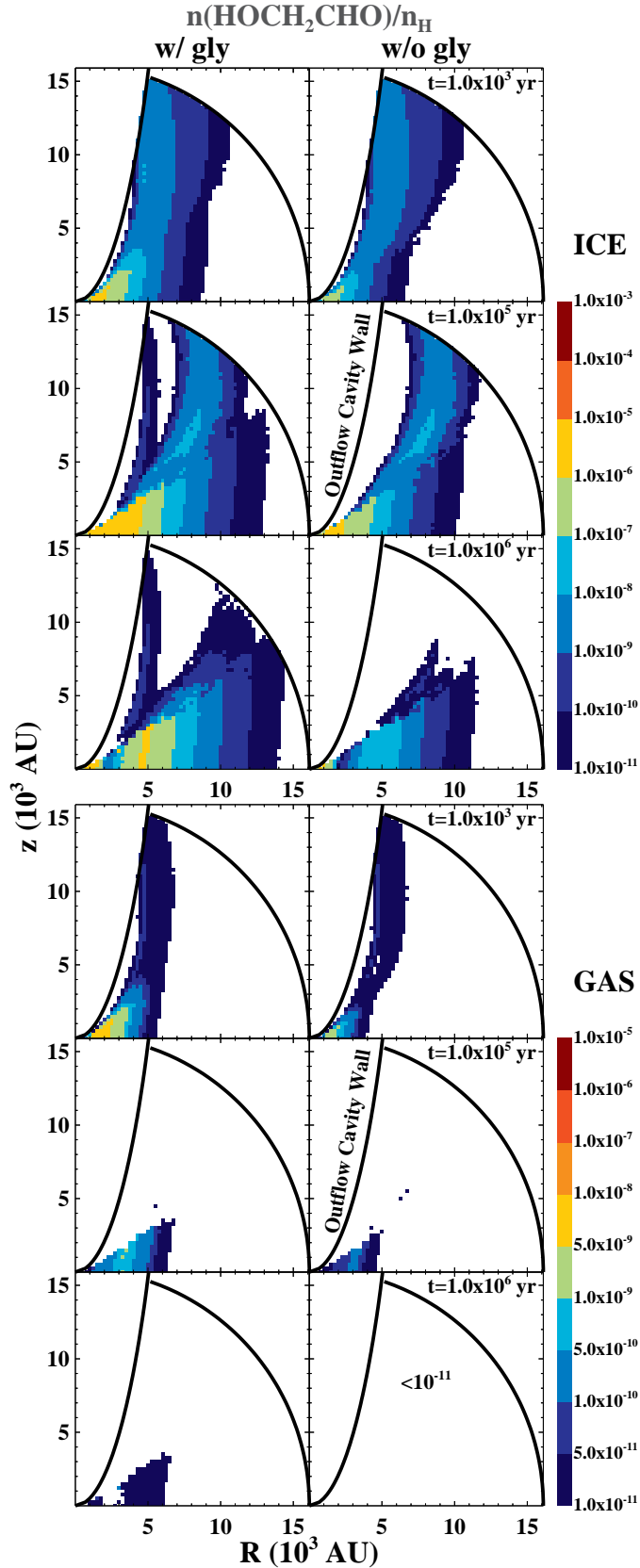}
 \caption{The abundance of glycolaldehyde in the solid (upper six panels) and gas (lower six panels) phases at three different time steps across the envelope-cavity system. The right column corresponds to the fiducial model. The left column is the model output upon the inclusion of the additional formation pathway via glyoxal. The outflow cavity wall is shown with a black curve. White cells correspond to either being outside of the area being considered or to having values outside of the range of the colour bar. The range of the gas colour bar is different from the range of that of the ice.}
 \label{fgr:GLY}
\end{figure*}

The chemistry explored in this work is limited by the reactions and the species included in our chemical network. In order to investigate this limitation, an additional formation pathway for glycolaldehyde is introduced, namely:
\[ {\rm H\overset{\bullet}{C}O_{ice} + H\overset{\bullet}{C}O_{ice} \rightarrow OCHCHO_{ice},} \]
\[ {\rm OCHCHO_{ice} \overset{H}{\rightarrow} 
\begin{aligned}
 & {\rm HO\overset{\bullet}{C}HCHO_{ice}} \\
 & {\rm \overset{\bullet}{O}CH_{2}CHO_{ice}} \\
\end{aligned}
 \overset{H}{\rightarrow} HOCH_{2}CHO_{ice}.} \]
This scheme was suggested by \citet{Woods2013} and experimentally proven for ices by \citet{Fedoseev2015}. In total, $68$ reactions are added to the existing chemical network and are detailed upon in Appendix~\ref{glynetwork}.

The same initial conditions are used as in the fiducial case (which also holds for all other test cases discussed), so that the dependence of the results purely on the inclusion of the additional pathway is explored. The abundance of gaseous and solid glycoladehyde increases by an order of magnitude across the system when the additional formation route is included, as Fig.~\ref{fgr:GLY} shows (the difference is larger in areas where the abundance of glycolaldehyde is very low ($n({\rm HOCH_{2}CHO_{ice}})/n_{\text{H}}<10^{-11}$), but that is not significant). This is consistent with the upper limit derived by \citet{Woods2013}, where destruction of glycolaldehyde and competitive routes are not considered. Adding an extra route for forming a molecule is expected to boost its abundance. The boost is larger in highly extincted and, in turn, cold ($T_{\text{dust}} \lesssim 20$~K) zones, where hydrogenation dominates over radical-radical associations. This enhancement also prolongs the lifetime of glycolaldehyde, and it is still found at appreciable ice abundances ($\sim 10^{-8}$) in the COM torus at $10^{6}$~yr, which was not the case previously. All other species show differences by a factor of a few or less (any modification of a chemical network results in non-linear effects on all species due to competing pathways, e.g., the inclusion of HCO~+~HCO competes with HCO~+~H, HCO~+~CH$_{3}$O, etc.), which is consistent with the findings of \citet{Fedoseev2015} using Monte Carlo models. This exercise shows that there may remain important routes to the formation of complex organics that have not yet been included in models.

Glycolaldehyde and methyl formate (and acetic acid) are isomers and may be intriguing probes of the physics and chemistry of a system. Recent observations have estimated the methyl formate to glycoladehyde ratio to be $\sim 13$ for IRAS16293-2422 \citep{Jorgensen2012, Coutens2015}, $\sim 12-20$ for NGC 1333-IRAS2A \citep{Coutens2015, Taquet2015} and $\sim 10$ for NGC 1333-IRAS4A \citep{Taquet2015}. Averaging over $10$ points in the COM torus (along $z=500$~AU and for $3756 \lesssim R \lesssim 6006$~AU) gives a ratio in the ice of $0.84$ fiducially and of $0.04$ after the route via glyoxal is introduced. The models indicate that there is more glycolaldehyde than methyl formate, while observations suggest the contrary. This may be explained by the fact that the observations probe predominantly the hot core regions, where gas-phase reactions become important \citep{Taquet2015}. They may drive chemistry that is different from that occuring in the solid phase. This may include gas-phase reactions with formic acid \citep{Taquet2015}, production of methoxy from OH+CH$_{3}$OH and hydrogen-abstraction reactions with F and Cl \citep{Balucani2015}.

\subsubsection{Stellar luminosity}

Young Class 0 and I protostars cover a range of luminosities from fractions of L$_{\sun}$ to more than an order of magnitude higher \citep{Evans2009, DunhamPPVI}. The dependence of these models on the luminosity is investigated by computing the chemistry for the case of $1~\text{L}_{\sun}$ and $15~\text{L}_{\sun}$. Our adopted luminosity of $35.7$~L$_{\sun}$ for NGC 1333-IRAS2A is on the higher side of the bulk of the low-mass protostars. Since the density is fixed, and both $F_{\text{FUV}}$ and the geometric dilution of the blackbody radiation scale with $L_{*}$ (see equation~\ref{tau}), the $A_{\text{V}}$ map remains unchanged for different luminosities. However, the number of FUV photons reaching a certain grid cell does change. Fig.~\ref{fgr:lum} in Appendix~\ref{suppfig} shows the difference in the dust temperature between the fiducial $35.7~\text{L}_{\sun}$ model and the two with lower luminosities. In the case of $15~\text{L}_{\sun}$, the envelope dust temperatures are cooler by at most $5$~K. For $1~\text{L}_{\sun}$, the differences are larger, and the cavity wall can be as much as $10-15$~K colder than in the original case.

Fig.~\ref{fgr:L} shows the methanol abundance in the case of $15~\text{L}_{\sun}$ and $1~\text{L}_{\sun}$ in comparison to the fiducial setup. For solid and gaseous methanol (with both phases displaying strong coupling), the abundances for $15~\text{L}_{\sun}$ and the fiducial run are very similar with differences predominantly within $1$ order of magnitude. Only small regions show an increase in methanol by $2-3$ orders of magnitude in the $15~\text{L}_{\sun}$ case. The enhancement along the cavity wall is explained by the shift of the temperature regime most favorable for grain-surface chemistry due to enhanced mobility prior to thermal desorption of radicals and the decrease of FUV photons making radical production inefficient. Other species (formic acid, methyl formate, acetaldehyde and dimethyl ether) predominantly show a decrease in abundance by an order of magnitude, leading to the increase in simpler species, like methanol and formaldehyde.

For the $1~\text{L}_{\sun}$ case (left column of Fig.~\ref{fgr:L}), the trends seen with $15~\text{L}_{\sun}$ are dramatized. The formation of complex organics is significantly impeded with gaseous and solid abundances of $\sim 10^{-11}$ or lower in the entire envelope. This results in an increase of solid, and in turn gaseous, methanol, by as much as $3$ orders of magnitude in a broad zone along the cavity. In conclusion, the original model with the highest luminosity of $35.7~\text{L}_{\sun}$ has been identified as most efficient for chemical complexity within this paradigm, while lower luminosities support stronger enhancement or retainment of simpler species, such as methanol. The morphology of the region, including the highlighting of the cavity wall, is preserved upon varying $L_{*}$, although the sizes of these regions change as do the peak abundances.

\subsubsection{Outflow cavity full opening angle}

\ctable[
 width = 0.4\textwidth,
 caption = {Outflow cavity parameters (additional model setups)}.,
 label = tbl:cavparam
 ]{@{\extracolsep{\fill}}llrr}{}{
 \hline
 Parameter & Units & Narrow & Wide \T\B\\
 \hline
 $a_{\text{cav}}$ & AU & $4.5 \times 10^{4}$ & $3.4 \times 10^{4}$\T\\
 $b_{\text{cav}}$ & AU & $2.0 \times 10^{3}$ & $8.0 \times 10^{3}$ \\
 $\alpha\left( z=1000 \text{AU} \right)$ & deg & $45~$ & $125~$ \\
 $\alpha\left( z=10~000 \text{AU} \right)$ & deg & $14~$ & $59~$\B\\
 \hline}

The shape of the outflow cavities and the cavity full opening angle set the flux of stellar photons penetrating the envelope, as mentioned in Section~\ref{physmod}. To test the importance of the full opening angle, two additional simulations have been executed, one with a narrower and another with a wider full opening angle (Table~\ref{tbl:cavparam}). Fig.~\ref{fgr:cav} in Appendix~\ref{suppfig} shows the difference in the dust temperature between the fiducial case and the two additional models, as well as the ratios in extinction. The temperature variations are largest in the zone that in one model is part of the cavity and not in the other, as well as in the cavity wall. The variations can exceed $20$~K close to the star, and are a result of the change in the density from the radial profile of the envelope to a fixed low value of the cavity. When comparing the three extinction maps, the pattern from narrow to fiducial to wide angle is that of an opening flower blossom. In the ratio plot, that is reflected by a large scale decrease in extinction of $7-25$ times in the cavity wall when going from the small cavity to the fiducial setup. When switching from the fiducial to the large cavity, a further decrease by a factor of $7-10$ is seen in an area immediately after the zone with the previous weakening. Furthermore, there is also a change in extinction in the zone that is switching from envelope to cavity area, as was the case with the dust temperature.

Fig.~\ref{fgr:C} shows the methanol abundance in the case of a wider and a narrower full opening angle in comparison to the fiducial cavity. The most prominent change is the enhancement by $2-3$ orders of magnitude $\sim5000$~AU away from the cavity wall, which is preceded by a decrease of $2-5$ orders of magnitude. This is due to the shift of the zone with efficient reactive desorption with the widening of the cavity. Other regions and other species predominantly show variations within an order of magnitude. The morphology of the system and highlighting of the cavity walls is preserved. Only the COM torus, still rich in complex organic species, is slightly reduced in size.

In the case of a small cavity (left column of Fig.~\ref{fgr:C}), complex organic molecules are only produced in a narrow strip in the immediate vicinity of the cavity wall, because in all other regions of the system, very few FUV photons penetrate making radical production inefficient. Methanol benefits from this and an enhancement in both solid and gas-phase abundances is seen for a much thicker layer along the cavity wall. In the fiducial setup, stellar FUV photons would actively photodissociate methanol in that zone. In essence, the full opening angle of a cavity determines primarily the amount of FUV radiation entering the envelope. The morphological structure of the system remains comparable, but enhancements in methanol and other complex organic molecules shift angularly with the migration of the regions with optimal temperature and extinction combinations.

%%%%%%%%%%%%%%%%%%%%%%%%%%%%%%%%%%%%%%%%%%%%%%%%%%%%%%%%%%%%%%%%%%%%%%%%%%%%%%%
\section{Discussion}
\label{discussion}

\begin{figure}
 \centering
 \includegraphics[keepaspectratio]{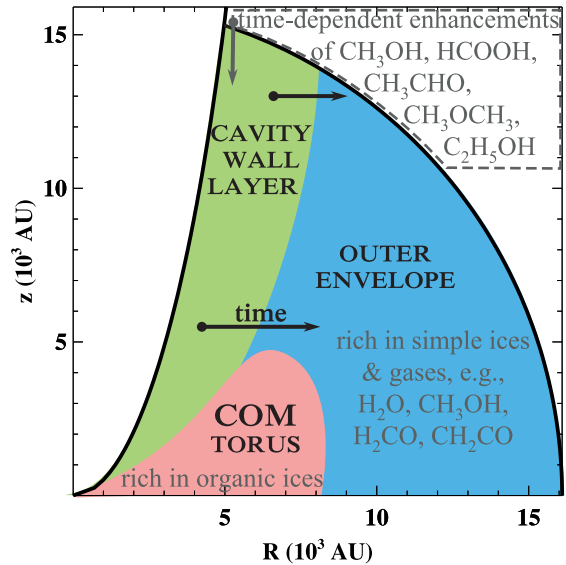}
 \caption{An illustration summarizing the key zones of the envelope-cavity system, and the major species and their phases therein. The motion of the cavity wall layer with time is indicated with two arrows. COM stands for complex organic molecules.}
 \label{fgr:morphcart}
\end{figure}

\begin{figure}
 \centering
 \includegraphics[keepaspectratio]{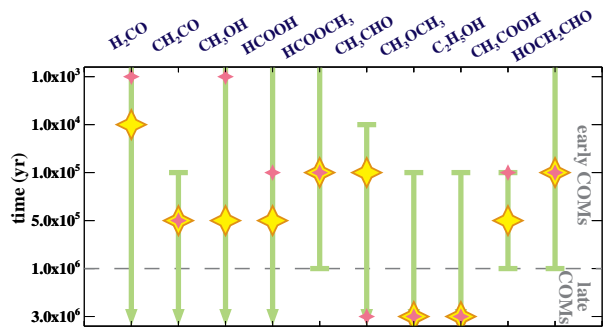}
 \caption{Lifetimes of complex organic ices in the COM torus (at $z=500$~AU). The green bars indicate when the abundance of the specified molecule is above $5 \times 10^{-8}$. The filled pink stars mark the time when the zone (in $R$) above this cut off value is maximized. The larger orange-yellow stars mark when the peak abundance is highest.}
 \label{fgr:COMs}
\end{figure}

\subsection{Morphology}

The abundance maps in Figs~\ref{fgr:CH3OH}-\ref{fgr:CH3OCH3} suggest an evolving chemical morphology of the envelope-cavity system. The representation in Fig.~\ref{fgr:morphcart} summarises all the information from the individual species in one global picture. First, there is the outflow cavity wall layer, which lights up in various species at different times. This region corresponds to where photodissociation by stellar UV photons leads preferentially to the formation of various large solid species via radical-radical associations on the grain. Reactive desorption stemming from many recombinations of photoproducts leads to an enhancement of methanol and complex organics in the gas phase. Acetaldehyde is an exception to this, due to an efficient gas-phase pathway (O~+~C$_{2}$H$_{5}$) leading to a large gas-phase abundance, but not in the solid phase. Such exceptional behaviour was also seen in and suggested by \citet{Codella2015}.

Secondly, there is the COM torus with the largest diversity of complex organic ices at abundances several orders of magnitude higher than initially injected into the system. The largest, most complex species (e.g., dimethyl ether in contrast to formic acid) appear later in time in the torus than in the cavity wall layer, but also survive much longer due to the mild radiation. This zone probes the time-scales of cool ($15-25$~K) quiescent ($A_{\text{V}} \sim 10$~mag) grain-surface chemistry with CR-induced FUV photons being the only source of radicals. In essence, the cavity wall layer is the scaled-up version of the torus, based on the scaling of the FUV radiation fields. A stronger stellar FUV radiation field builds complex organics faster and in a larger zone, but for a short period of prosperity. A weaker CR-induced FUV radiation field builds complexity more slowly, in a smaller zone that is shielded from powerful dissociating stellar radiation. Both require cool dust temperatures (the distinguishing feature from the prestellar phase), and both eventually destroy complexity via photodissociation.

Finally, the rest of the envelope is where only relatively simple ices and gases reside. This pertains to species like water, methanol, formaldehyde, and ketene among others. Few changes happen in this zone compared with the initial conditions, because the temperature is low ($\lesssim 15$~K) and shielding is high ($A_{\text{V}} \gtrsim 3$~mag). The chemical composition of the outer region remains similar to the prestellar phase.

Several important lessons can be learned from the abundance maps, besides the chemical morphology of the system. As is seen for the COM torus, gases do not always reflect the composition of the ices. Effective non-thermal desorption mechanisms are needed to couple the two phases in the predominantly cold ($<100$~K) envelope. Furthermore, there may be exceptional species, like acetaldehyde, that have an efficient gas-phase route at particular physical conditions (like that of the cavity wall layer), which would result in an inferred ice abundance that is several orders of magnitude too high. Observations of solids are needed to accurately constrain the icy content, which remains a challenging task for solid complex organic molecules. Some species with isolated features in the $5-10$~$\mu$m range can be searched for with future missions like the \textit{James Webb Space Telescope} (\textit{JWST}).

\subsection{Comparison with observations}

From Fig.~\ref{fgr:CH3OH} it can be seen that gaseous methanol illuminates the cavity wall as early as $10^{3}$~yr with abundances of $\sim 10^{-8}$. Formic acid, acetaldehyde, dimethyl ether and ethanol reach gas-phase abundances of $\sim 10^{-9}$, $10^{-8}$, $10^{-9}$ and $10^{-9}$, respectively, in the cavity wall layer.  These values are in rough agreement with the column densities derived by \citet{Arce2008} for formic acid and ethanol at the B1 position ($\sim 10^{13}-10^{14}$~cm$^{-2}$ giving abundances $\sim 10^{-10}-10^{-9}$~cm$^{-3}$), as well as those for acetaldehyde and dimethyl ether ($\sim 10^{13}$~cm$^{-2}$) from \citet{Oberg2011Serpens} towards Serpens. This implies that thermal desorption (for $T_{\text{dust}} \gtrsim 100$~K) and sputtering with the passage of shocks along cavity walls do not need to be invoked when interpreting observations on these scales (several thousand AU). These processes are more likely to be efficient on smaller (several hundred AU) scales and shorter time-scales ($\sim 10^{3}$~yr).  The enhancements seen in the cavity wall layer rely heavily on reactive desorption, which dominates over much less efficient direct photodesorption. This highlights the need for further laboratory studies quantifying the efficacy of reactive desorption as a mechanism for releasing complex organic ices into the gas phase at low temperatures. If it is inefficient, then shocks may once again be necessary to explain observations. Methyl formate does not follow the observed abundances of \citet{Arce2008} and \citet{Oberg2011Serpens} in our models, and in fact, is only efficiently made in the COM torus.

By design of the chemical network used in this work, methanol plays a central role in the synthesis of larger, more complex species. However, it appears not only as a precursor, but also as a descendent (or postcursor), once CR-induced FUV photons dominate and destroy complexity at longer time-scales ($\gtrsim 3 \times 10^{6}$~yr). This scenario may be challenged as more atom addition reactions are included into the chemical network.

Our models predict the COM torus -- a solid-state sweet spot, which has not been seen observationally. The abundances in this zone may still change due to influences of the protoplanetary disk and other dynamical effects (Section~\ref{epiacc}). For NGC 7538 IRS9 -- a massive YSO, \citet{Oberg2013} reported a change in chemistry within $8000$~AU of the protostar, which potentially links with efficient grain-surface chemistry. It remains unclear whether a high-mass source would be associated with more efficient pathways towards enriching the gas phase from the ices rich in complex organics, in comparison to the low-mass case modelled in this work. Although, recent ALMA observations towards the high-mass IRAS16547-4247 by \citet{Higuchi2015} suggest that methanol emission follows an hourglass morphology.

\subsection{Lifetimes of complex organics}

From the abundance maps it was seen that species peak at different times. Motivated by this, the ordering of complex organic species in time is shown in Fig.~\ref{fgr:COMs} as they appear in the COM torus (a slice at $z=500$~AU is assumed to be representative, which covers $R \in \left[ 1006, 16~100 \right]$~AU). The green bars show when the abundance of an ice is above $5 \times 10^{-8}$, i.e., when the solid phase is enhanced. The time of maximal spatial extent of this enhancement along $R$ is marked with a filled pink star. When the peak ice abundance is reached, this is indicated with a larger orange-yellow star. The two do not necessarily overlap for all species considered. As was hinted earlier, Figs~\ref{fgr:HCOOCH3},~\ref{fgr:CH3CHO},~\ref{fgr:CH3OCH3} and~\ref{fgr:COMs} do not support the partition of methyl formate with acetaldehyde into the group of `cold' organics, contrasting dimethyl ether as a `hot' species. Instead, acetaldehyde appears to have more in common with the morphology and lifetime of dimethyl ether. Yet, all three are formed in the COM torus at cool temperatures. A different organization of complex organic molecules by their lifetimes and times of peak abundance is attempted in Fig.~\ref{fgr:COMs}. This would imply that species like formaldehyde, ketene, methanol, formic acid, methyl formate, acetic acid and glycolaldehyde are `early' types; while acetaldehyde, dimethyl ether and ethanol are `late' types. Fig.~\ref{fgr:COMs} indicates a chemical pattern that is consistent with the chemistry invoked in our model. Ketene, dimethyl ether and ethanol reach abundances $> 5 \times 10^{-8}$ only at $10^{5}$~yr, while acetic acid abundances do not stay high outside of the $10^{5}-10^{6}$~yr range. This implies that simple ices, like methanol and ketene, are converted into more complex organic ices with the evolutionary stage of the source (Class 0 to Class I). \citet{Oberg2013} proposed that ketene, methanol and acetaldehyde belong to `zeroth-generation' molecules, which is supported in this work for the former two species, but not for the latter (although the abundance of acetaldehyde peaks earlier than the maximization of the spatial extent of its enhancement).

\subsection{Episodic accretion and dynamics}
\label{epiacc}

A potential caveat to the modelling presented thus far is episodic accretion. \citet{Visser2012episodic} and \citet{Visser2015} performed the first studies towards understanding the chemical implications of an accretion burst. The authors estimated that the effects should remain observable up to $10^{4}$~yr after such an event and that there should be an excess of gaseous CO. A factor $10$ increase in stellar luminosity (typical for a moderately strong burst) is expected to push the $100$~K radius in the envelope out from $\sim 38$ to $\sim 114$~AU and the $20$~K radius from $\sim 3300$ to $\sim 10~000$~AU for our setup \citep{Visser2012episodic}. Thus strong heating of the dust leading to thermal desorption of all species, in turn transferring the complexity from the solid phase into the gas, only occurs for the inner envelope and does not influence the large scales studied here. However, if episodic accretion is accompanied by sufficiently strong FUV irradiation (stronger than the typical interstellar UV field), which would photodissociate complex organic molecules, complexity may be reset for regions of the system. Alternatively, if FUV irradiation and/or dust heating are mild, then complexity may be enhanced, especially now that the $20$~K radius lies much further out. The effects of episodic accretion must strongly depend on the exact physical properties of such an event, which is beyond the scope of this paper.

In addition, the static model used in this work neglects the dynamical effects of the system. On smaller scales, such as the inner $\sim 500$~AU studied in \citet{Drozdovskaya2014}, dynamics definitely play a crucial role due to the proximity to the growing protostar and the expanding protoplanetary disc, which is also where material moves through regions with high temperatures. However, on much larger scales of several thousand AU that are focused on in this work, a static model may be sufficient as the physical conditions have a shallower gradient with distance through the envelope. Furthermore, \citet{Jorgensen2002} and \citet{Kristensen2012} suggest that although Class 0 envelopes are infalling, by the Class I phase some envelopes are already showing expansion.

%%%%%%%%%%%%%%%%%%%%%%%%%%%%%%%%%%%%%%%%%%%%%%%%%%%%%%%%%%%%%%%%%%%%%%%%%%%%%%%
\section{Conclusions}
\label{conclusions}

The envelopes encompassing young low-mass protostars are successively eroded by bipolar outflows, resulting in cavities that grow with time. These outflow cavities allow additional photons to penetrate the envelope, thus enhancing the heating of the dust and the irradiation of the material therein. Various observations (e.g., \citealt{Arce2008, Oberg2011}) have hinted that this results in unique morphological features and gas-phase detections of complex organic molecules. This paper describes a 2D physicochemical model of such an envelope-cavity system (excluding a traditional hot corino), including wavelength-dependent radiative transfer calculations and a comprehensive gas-grain chemical network. The main results of our modelling are:

\begin{enumerate}
	\item An envelope-cavity system has three distinct regions. Firstly, the cavity wall layer, which displays time-dependent enhancements in solid and gaseous abundances of methanol, formic acid, acetaldehyde, dimethyl ether and ethanol. Secondly, closer to the star, a `COM torus' -- a zone rich in complex organic ices. Thirdly, the remaining outer envelope, which is predominantly comprised of simpler ices and gases, like water, methanol, formaldehyde and ketene.
	\item Gases do not always reflect the ice composition -- the COM torus is poor in gaseous complex organics, while having the highest solid abundances of the entire system. The division between the two phases depends on the relative efficiencies of non-thermal desorption mechanisms in different regions.
	\item Complex organic ices peak in the COM torus at different times suggesting unique molecular lifetimes.
	\item The gas-phase enhancements along cavity walls in select molecules are consistent with observations. In strongly irradiated regions ($A_{\text{V}}<3$~mag), such as the cavity wall layer, photodissociation in the solid phase is frequent. Subsequent recombination of the photoproducts leads to frequent reactive desorption. Although the poorly constrained efficiency is assumed to be a mere $1$ per cent, the high frequency of reactive desorption leads to gas-phase enhancements of several orders of magnitude. Direct photodesorption (using the laboratory constrained yield of $10^{-3}$ molecules per photon) is found to be inefficient in comparison, indicating the need for experimental quantification of the rates for reactive desorption.
	\item Photodissociation directly in the solid phase produces radicals in the icy mantle, the absence of which impedes the formation of complex organic molecules. For efficiency, a sufficient number of FUV photons needs to penetrate the envelope and elevated cool dust temperatures enable radical mobility on the grain surface.
	\item Consequently, a high stellar luminosity ($\sim 35$~L$_{\sun}$) favours chemical complexity. A low stellar luminosity ($\sim 1$~L$_{\sun}$) results in negligible complex organic abundances ($<10^{-11}$).
	\item Also, a sufficiently wide cavity (for example, $\alpha \left( z=10~000\text{AU} \right)=45^{\circ}$) is needed to directly irradiate the envelope.	
\end{enumerate}

This work has shown that low-mass protostars have gas-phase enhancements in complex organic molecules when outflow cavities are present, as well as hidden zones that are rich in complex organic ices. Future research will explore the intriguing extrapolation of this to the high-mass case, especially in light of recent ALMA observations of methanol emission with an hourglass morphology towards IRAS16547-4247 \citep{Higuchi2015}.

%%%%%%%%%%%%%%%%%%%%%%%%%%%%%%%%%%%%%%%%%%%%%%%%%%%%%%%%%%%%%%%%%%%%%%%%%%%%%%%
\section{Acknowledgements}
\label{acknowledgements}

This work is supported by a Huygens fellowship from Leiden University, by the European Union A-ERC grant 291141 CHEMPLAN, by the Netherlands Research School for Astronomy (NOVA) and by a Royal Netherlands Academy of Arts and Sciences (KNAW) professor prize. C.W. acknowledges support from the Netherlands Organisation for Scientific Research (NWO, program number 639.041.335).

%%%%%%%%%%%%%%%%%%%%%%%%%%%%%%%%%%%%%%%%%%%%%%%%%%%%%%%%%%%%%%%%%%%%%%%%%%%%%%%
\bibliographystyle{mn2e}
\bibliography{mybib} % mybib.bib file

%%%%%%%%%%%%%%%%%%%%%%%%%%%%%%%%%%%%%%%%%%%%%%%%%%%%%%%%%%%%%%%%%%%%%%%%%%%%%%%
\newpage
\appendix

\section[]{Additional abundance maps}
\label{addmaps}

In addition to the five sets of abundance maps shown in the main text of the paper, another five molecules are presented here for completeness.

\begin{figure}
 \centering
 \includegraphics[keepaspectratio]{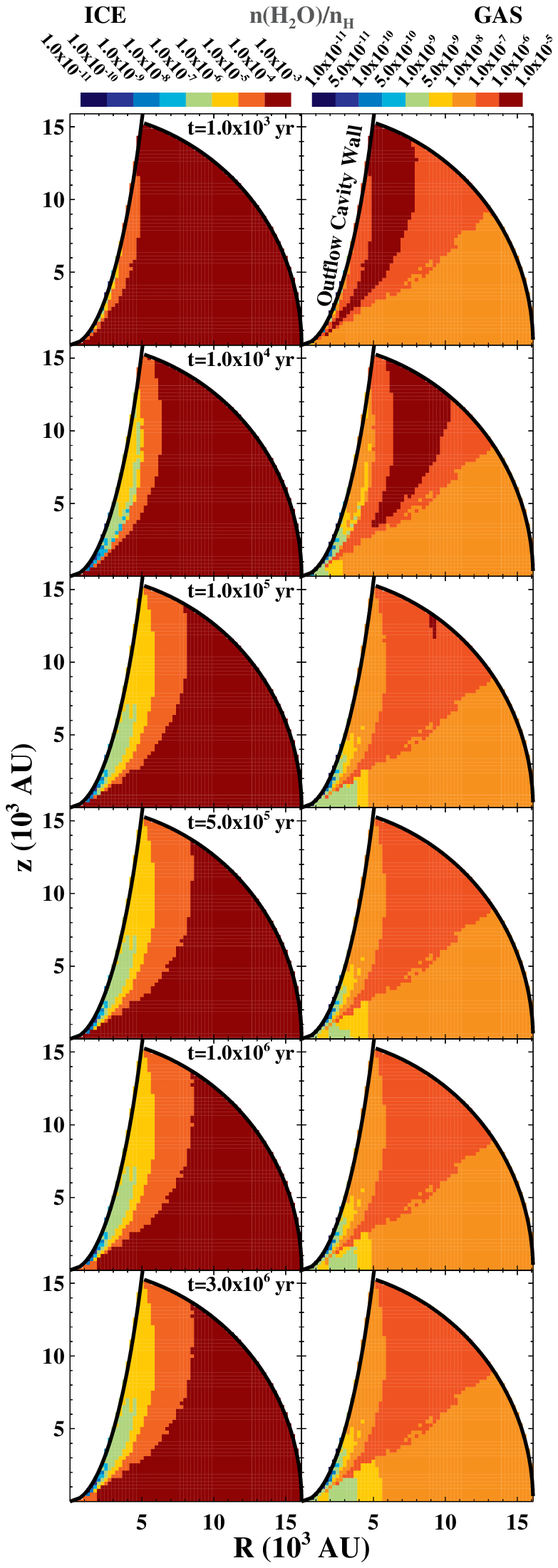}
 \caption{Same as Fig.~\ref{fgr:CH3OH}, but for water.}
 \label{fgr:H2O}
\end{figure}

\begin{figure}
 \centering
 \includegraphics[keepaspectratio]{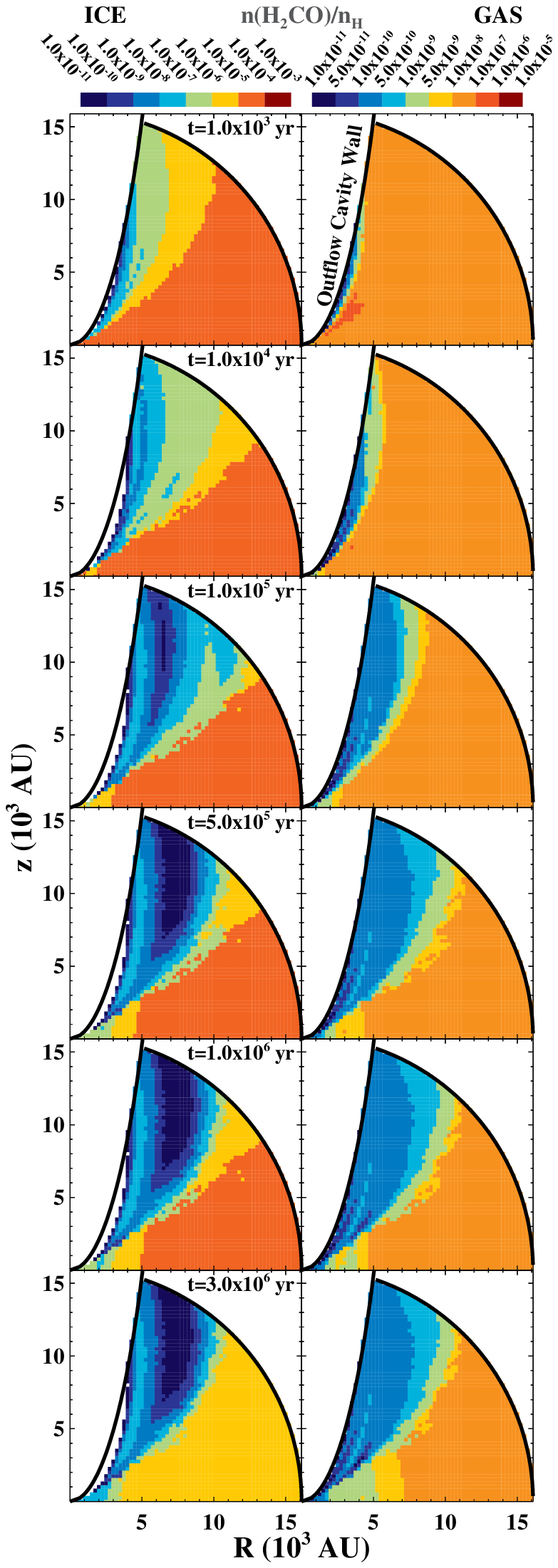}
 \caption{Same as Fig.~\ref{fgr:CH3OH}, but for formaldehyde.}
 \label{fgr:H2CO}
\end{figure}

\begin{figure}
 \centering
 \includegraphics[keepaspectratio]{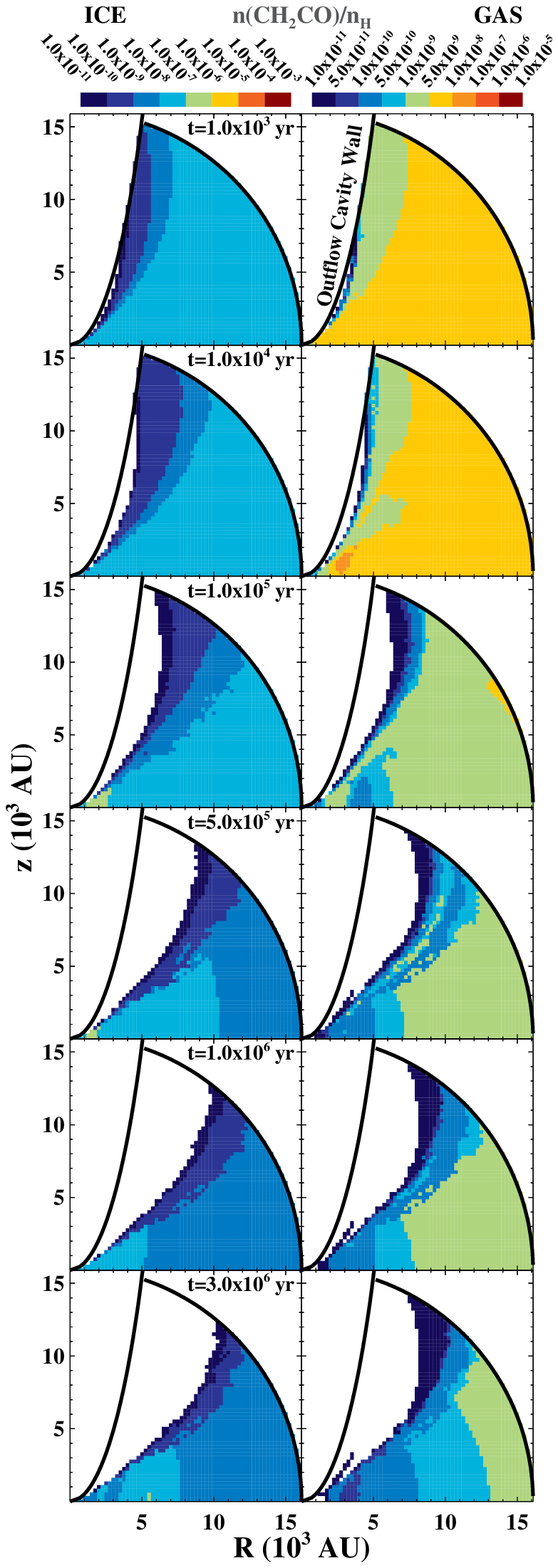}
 \caption{Same as Fig.~\ref{fgr:CH3OH}, but for ketene.}
 \label{fgr:CH2CO}
\end{figure}

\begin{figure}
 \centering
 \includegraphics[keepaspectratio]{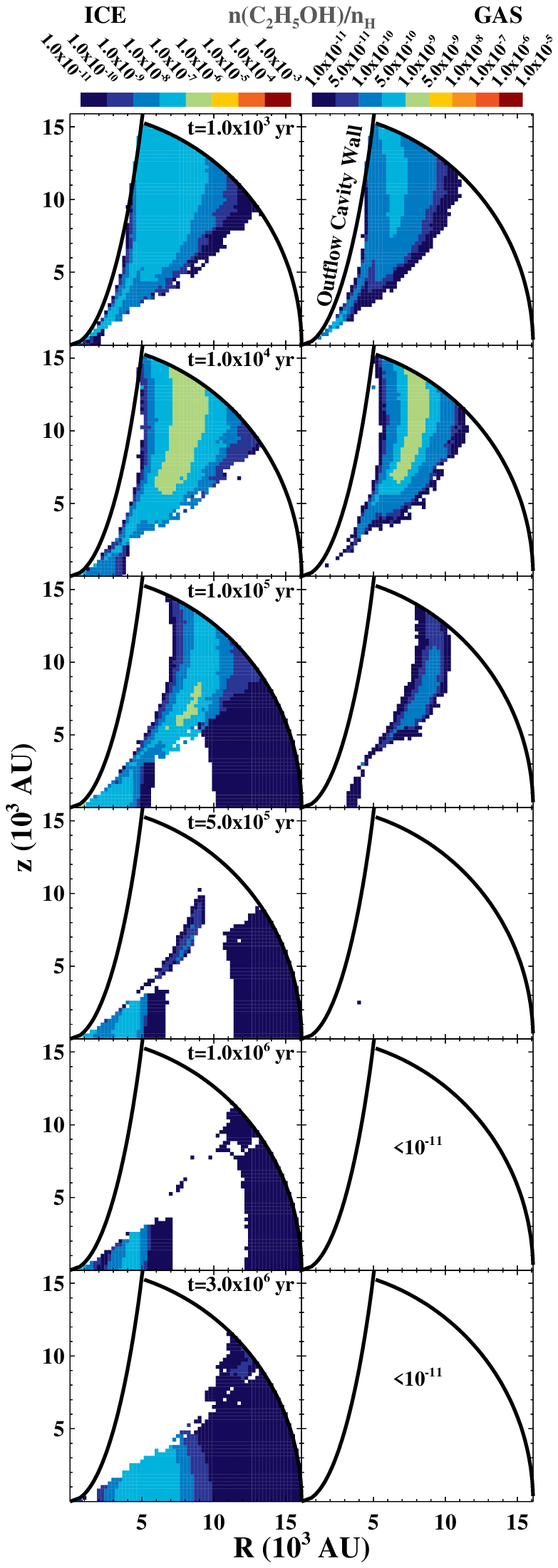}
 \caption{Same as Fig.~\ref{fgr:CH3OH}, but for ethanol.}
 \label{fgr:C2H5OH}
\end{figure}

\begin{figure}
 \centering
 \includegraphics[keepaspectratio]{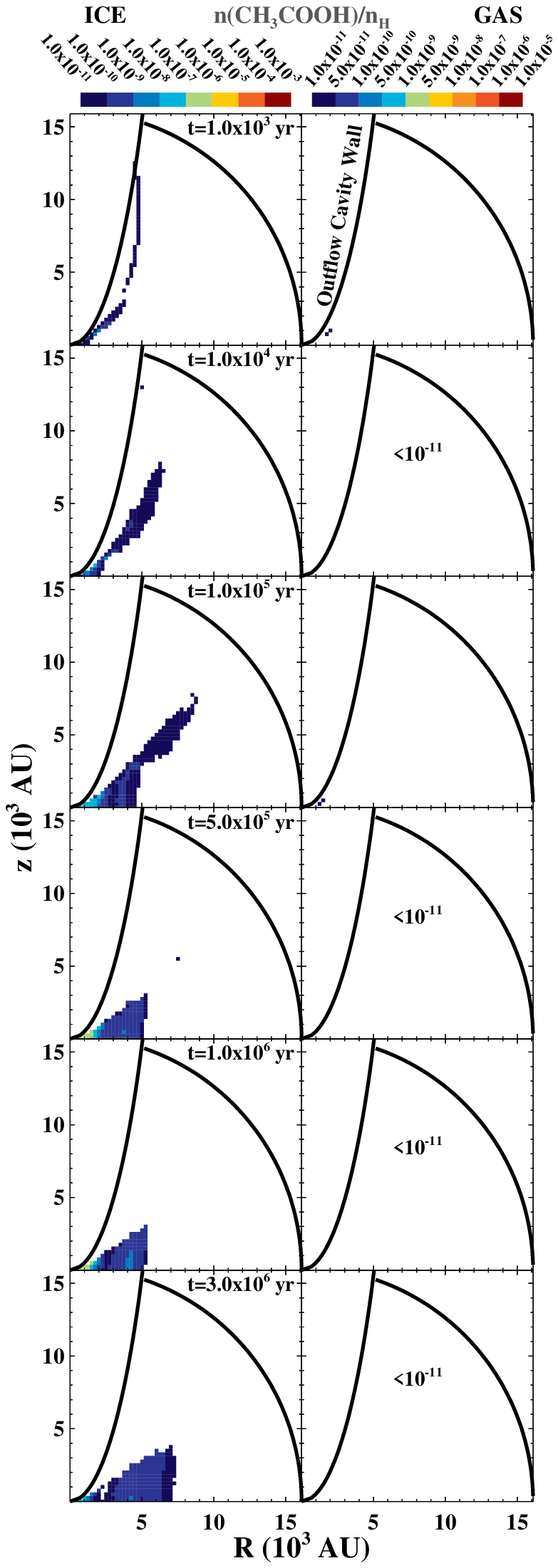}
 \caption{Same as Fig.~\ref{fgr:CH3OH}, but for acetic acid.}
 \label{fgr:CH3COOH}
\end{figure}

\newpage
\section[]{Details of the route to glycolaldehyde via glyoxal}
\label{glynetwork}
The details of the added route to glycolaldehyde via glyoxal are described in this appendix. In this work, the route is only added for the solid phase (unlike in \citealt{Woods2013}), since three-body gas phase reactions require a radiative association mechanism, which is slow by nature. Associations of radicals are assumed to be barrierless, which includes the reaction between the two formyl radicals (${\rm H\overset{\bullet}{C}O}$) and the second hydrogenation. Hydrogenation of glyoxal (OCHCHO) has an adopted activation energy ($E_{\text{A}}$) of $\sim 1100$~K, based on quantum chemical calculations of \citet{Woods2013}.

The chemical network used is thereby expanded by $10$ new species: gaseous and solid glyoxal, the cation of glyoxal (OCHCHO$^{+}$), the protonated glyoxal (OCHCH$_{2}$O$^{+}$), the gaseous and solid two intermediate species (${\rm HO\overset{\bullet}{C}HCHO}$, ${\rm \overset{\bullet}{O}CH_{2}CHO}$), a joint cation (C$_{2}$O$_{2}$H$_{3}^{+}$), and a joint protonated form of the intermediates (C$_{2}$O$_{2}$H$_{4}^{+}$). The gas-phase chemistry of glyoxal is extracted from the OSU network \citep{Garrod2008}. The gas-phase chemistry of the two intermediate species is assumed to be similar to that of other large hydrogenated radicals and glyoxal itself, from which the gas-phase reaction rate coefficients are adopted. The dissociative reaction channels with He$^{+}$ and photodissociation channels are assumed to be 2HCO~+~H for ${\rm HO\overset{\bullet}{C}HCHO}$ and H$_{2}$CO~+~HCO for ${\rm \overset{\bullet}{O}CH_{2}CHO}$. Grain-surface chemistry of glyoxal and the two intermediates is not expanded further than the scheme provided above, with the exception of one competing reaction, namely:
\[ {\rm OCHCHO_{ice} \overset{H}{\rightarrow} H_{2}CO_{ice} + HCO_{ice}}, \]
which is included in the OSU network (again $E_{\text{A}}=1108$~K). The binding energy of glyoxal is taken to be $3200$~K (OSU network, \citealt{Garrod2008}) and for the two intermediates calculated according to the prescription in \citet{Garrod2008} to be $E_{\text{des}} ( {\rm HO\overset{\bullet}{C}HCHO_{ice}} )=6080$~K and $E_{\text{des}} ( {\rm \overset{\bullet}{O}CH_{2}CHO} )=3800$~K (which accounts for the former molecule being able to H-bond due to the presence of an OH functional group).

\section[]{Supporting parameter study figures}
\label{suppfig}

\begin{figure*}
 \centering
 \includegraphics[keepaspectratio]{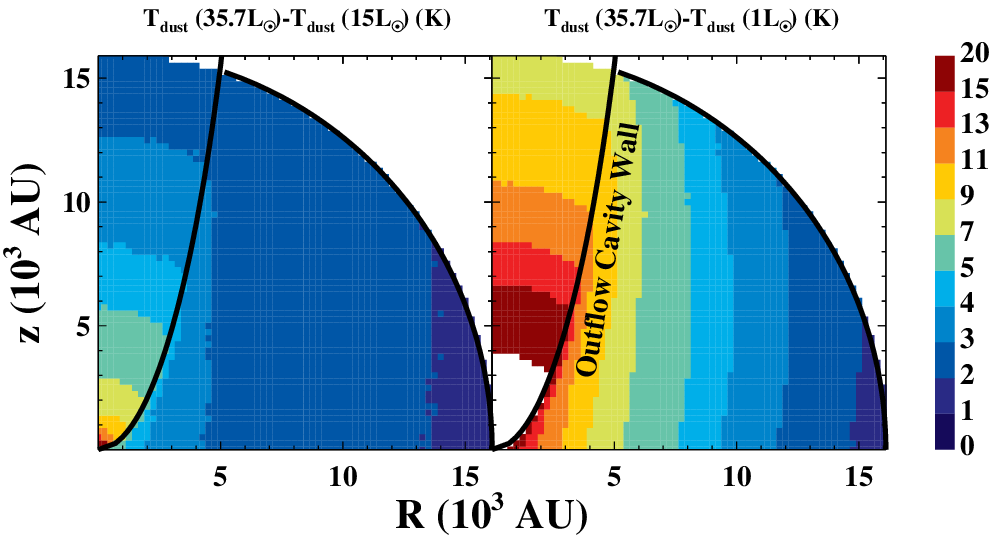}
 \caption{The difference in the dust temperature between the fiducial, $35.7~\text{L}_{\sun}$ model, and the two with smaller luminosities ($15~\text{L}_{\sun}$ and $1~\text{L}_{\sun}$). The outflow cavity wall is shown with a black curve.}
 \label{fgr:lum}
\end{figure*}

\begin{figure*}
 \centering
 \includegraphics[keepaspectratio]{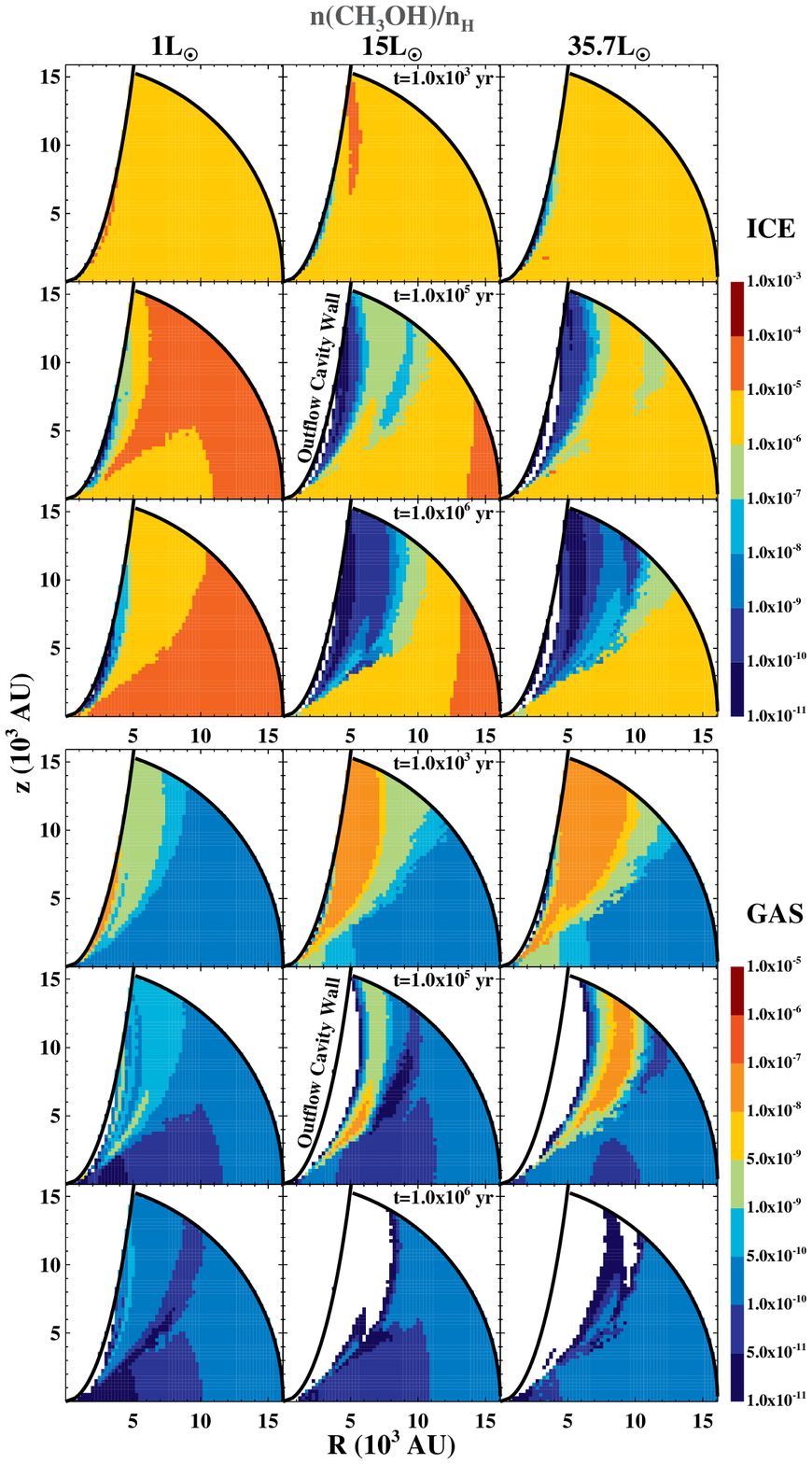}
 \caption{The abundance of methanol in the solid (upper nine figures) and gas (lower nine figures) phases at three different time steps across the envelope-cavity system. The left, middle and right columns correspond to the cases of $1~\text{L}_{\sun}$, $15~\text{L}_{\sun}$ and $35.7~\text{L}_{\sun}$ (the fiducial model), respectively. The outflow cavity wall is shown with a black curve. White cells correspond to either being outside of the area being considered or to having values outside of the range of the colour bar.}
 \label{fgr:L}
\end{figure*}

\begin{figure*}
 \centering
 \includegraphics[keepaspectratio]{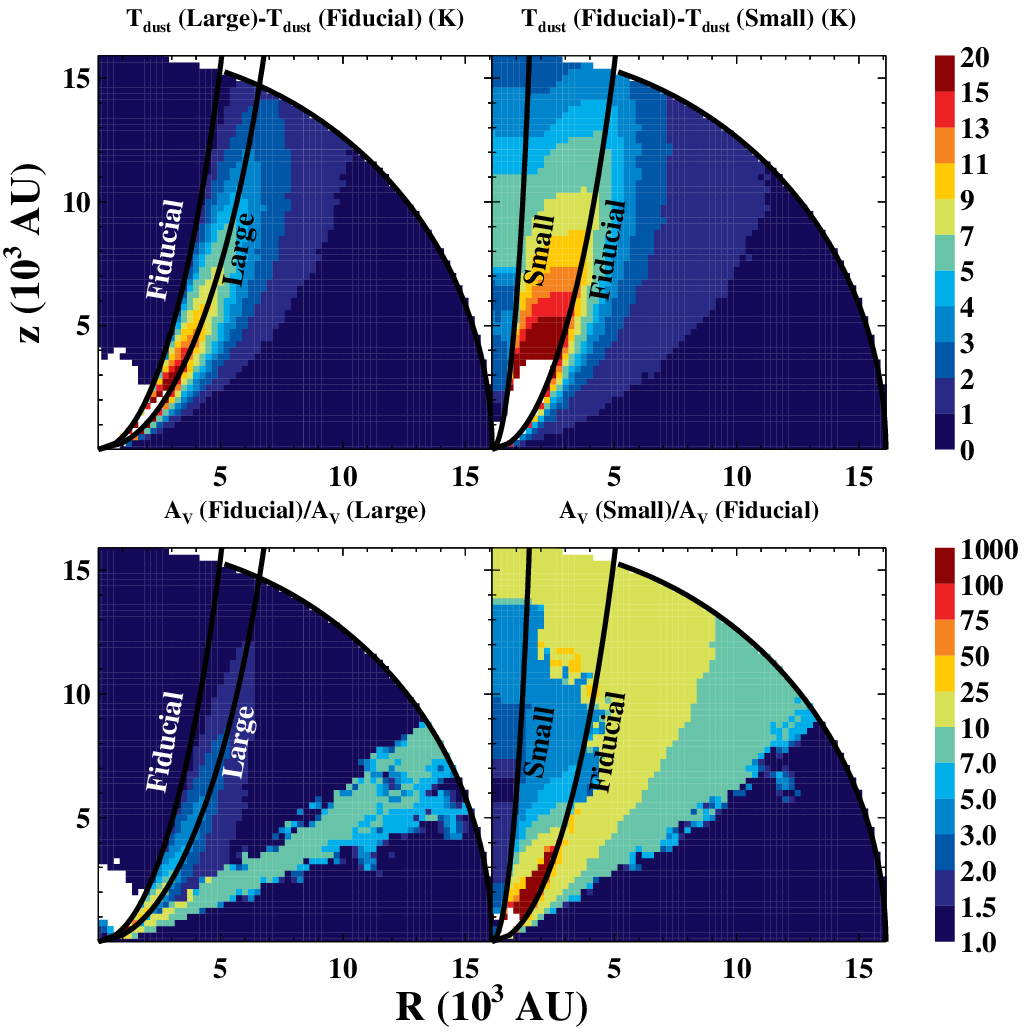}
 \caption{The difference in the dust temperature between the fiducial model and the two additional models, one with a larger cavity (upper left panel) and another with a smaller cavity (upper right panel), as well as the ratios in extinction (lower left and right panels, respectively). The different outflow cavity walls are shown and labeled.}
 \label{fgr:cav}
\end{figure*}

\begin{figure*}
 \centering
 \includegraphics[keepaspectratio]{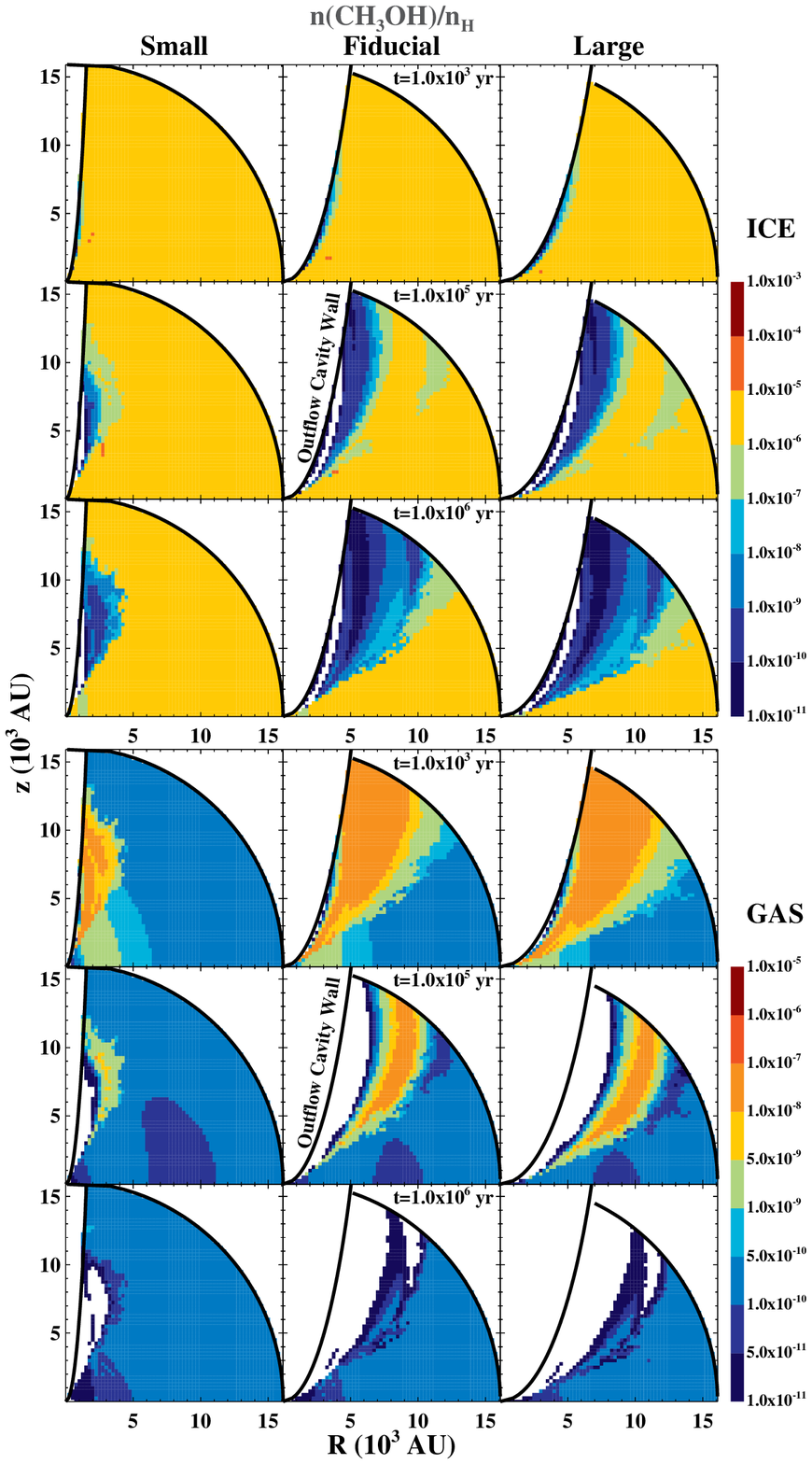}
 \caption{The abundance of methanol in the solid (upper nine figures) and gas (lower nine figures) phases at three different time steps across the envelope-cavity system. The left, middle and right columns correspond to the cases of a small, fiducial and large cavities, respectively. The outflow cavity wall is shown with a black curve. White cells correspond to either being outside of the area being considered or to having values outside of the range of the colour bar.}
 \label{fgr:C}
\end{figure*}

\newpage

%%%%%%%%%%%%%%%%%%%%%%%%%%%%%%%%%%%%%%%%%%%%%%%%%%%%%%%%%%%%%%%%%%%%%%%%%%%%%%%
\bsp % ``This paper has been produced using the ...''

\label{lastpage}

\end{document}